\documentclass[12pt]{article}
\usepackage{amsfonts}
\usepackage{amsmath}
\usepackage{theorem}
\usepackage{enumerate}
\usepackage{latexsym}
\usepackage{graphicx}

\def\a{\alpha}

\def\d{\delta}

\def\l{\lambda}

\def\e{\epsilon}

\newtheorem{teo}{Theorem}[section]
\newtheorem{lem}{Lemma}[section]
\newtheorem{tve}{Proposition}[section]
\newtheorem{nas}{Corollary}[section]

\title{\bf  Classical double, $R$-operators and negative flows of  integrable hierarchies.}
\author{B. Dubrovin $^{1,2}$, T. Skrypnyk $^{1,3}$\\
{\small $^1$ International School for Advanced Studies,} {\small
SISSA, via Bonomea 265, 34136 Trieste, Italy}
\\{\small $^2$ Steklov Mathematical Institute, Moscow, Russia}
\\
{\small  $^3$ Bogolyubov Institute for Theoretical Physics,
Metrolohichna str.14-b, Kiev, 03680, Ukraine }}
\date{ }

\begin{document}
\thispagestyle{empty}
 \maketitle
 \begin {abstract}
Using classical  double $ {\mathcal{G}}$ of a  Lie
algebra $ {\mathfrak{g}}$ equipped with the classical
$R$-operator we define two sets of mutually commuting functions
with respect to the initial Lie--Poisson bracket  on
$ {\mathfrak{g}}^*$ and its extensions. We consider in
details examples of the Lie algebras $ {\mathfrak{g}}$
with the ``Adler--Kostant--Symes" $R$-operators and the corresponding
two sets of mutually commuting functions. Using the constructed
commutative hamiltonian flows on different extensions of
$ {\mathfrak{g}}$ we obtain zero-curvature equations with
$ {\mathfrak{g}}$-valued $U$-$V$ pairs. Among such the
equations are so-called ``negative flows'' of soliton hierarchies.
We illlustrate our approach by examples of abelian and non-abelian
Toda field equations.
\end{abstract}


\section{Introduction}
The theory of  hierarchies of integrable partial differential equations is based on the possibility to represent each of the
equations of the hierarchy in the so-called zero-curvature form
$$
U_t-V_x+[U,V]=0,
$$
with the corresponding $U$-$V$-pair
taking values in some infinite-dimensional Lie algebra $ {\mathfrak{g}}$.

There are several approaches to a construction of zero-curvature
equations starting from Lie algebras $ {\mathfrak{g}}$.
One of the most simple and straightforward of them  is the
approach of \cite{FNR}, \cite{New} that interprets zero-curvature
equations as a compatibility condition of two auxiliary
Lax (Euler--Arnold) flows. The commutativity of these Lax flows is
guaranteed by the Poisson-commutativity of the corresponding
hamiltonians. In this approach the elements $U$ and $V$ from
zero-curvature equations coincide with the algebra-valued
gradients of the commuting hamiltonians constructed with the help
of the
 Adler--Kostant--Symes scheme.
In more details, these hamiltonians coincide with the
 restrictions of  Casimir
 functions of $ {\mathfrak{g}}$ onto the dual spaces to subalgebras
 $\mathfrak{g}_{\pm}$, were $\mathfrak{g}=\mathfrak{g}_+ + \mathfrak{g}_-$.
Such an approach permits one  to
construct two types
of integrable equations associated with the Lie algebra $\mathfrak{g}$, namely, integrable equations with the
elements  $U$ and $V$ belonging to the same Lie subalgebras
$\mathfrak{g}_{\pm}$.

However the approach of \cite{New}
does not cover all known integrable equations. In particular, it does not work for integrable equations (sometimes called \emph{negative flows} of integrable hierarchies)
possessing $U$-$V$-pairs with $U$-operator belonging to
$\mathfrak{g}_{+}$ and $V$-operator belonging to
$\mathfrak{g}_{-}$. In the paper \cite{SkrTMP1} such
equations were included into the general scheme by showing
that the  restrictions of  Casimir
 functions of $\mathfrak{g}$ onto the dual spaces to subalgebras
 $\mathfrak{g}_{+}$ and $\mathfrak{g}_{-}$  commute not only inside each group but also between the groups.
This permits to construct  negative flows of integrable
hierarchies as a consequence of commutativity of Lax flows
generated by ``positive" and ``negative" hamiltonians. In the papers
\cite{SkrTMP2}, \cite{SkrSIGMA}  it was proposed a generalization of
the above scheme onto the case of Lie algebras
$\mathfrak{g}$ possessing a general classical $R$
operator not always connected with the decomposition
$\mathfrak{g}=\mathfrak{g}_+ +
\mathfrak{g}_-$ (i.e, not always connected with the
Adler--Kostant--Symes scheme). It was shown that the restrictions of
Casimir
 functions of $\mathfrak{g}$ onto the subalgebras
 $\mathfrak{g}_{R_{\pm}}$, where $\mathfrak{g}_{R_{\pm}}=\mathrm{Im} R_{\pm}$
  commute not only inside each group but also between the groups.
  This observation  permits one to obtain two sets of mutually
commuting functions on $\mathfrak{g}^*$ and three
types of zero-curvature equations, in particular those
corresponding to negative flows of soliton hierarchies
\cite{SkrSIGMA}. Observe that the corresponding commutativity
does not follow from the standard $R$-matrix scheme \cite{ST} on
$\mathfrak{g}$.

It turns out that the scheme proposed in the papers
\cite{SkrTMP2}, \cite{SkrSIGMA}  is still not  the most general
approach towards a generation of commutative flows on
$\mathfrak{g}$ and, hence,  not the most general
approach to the construction of soliton hierarchies with
$\mathfrak{g}$-valued $U$-$V$ pairs. In particular, it
does not include the infinite-component Toda hierarchy and does
not produce the corresponding auxiliary Lax equations
\cite{Ueno}.

In the present paper we propose a more general method of
constructing commutative flows and zero-curvature equations
with  $\mathfrak{g}$-valued $U$-$V$ pairs. For this
purpose we consider commutative flows not on
$\mathfrak{g}^*$ but on $\mathcal{G}^*$,
where $\mathcal{G}$ is a classical double of
$\mathfrak{g}$. We utilize the fact that a classical
$R$-operator on
$\mathfrak{g}$ induces a natural $R$-operator $\mathcal{R}$
on $\mathcal{G}= \mathfrak{g}\oplus
\mathfrak{g}$ \cite{RST2}. This $R$-operator $\mathcal{R}$ on $\mathcal{G}$ proves to be \cite{RST2}
always of the Adler--Kostant--Symes type, regardless the form of the
original operator $R$ on $\mathfrak{g}$. From this it
follows that $\mathcal{G}_{\mathcal{R}}=
\mathcal{G}_{\mathcal{R}_{+}}\ominus
\mathcal{G}_{\mathcal{R}_{-}}$, where algebra
$\mathcal{G}_{\mathcal{R}}$ is a linear space
$\mathcal{G}$ equipped with so-called
$\mathcal{R}$-bracket \cite{ST}. Moreover it turns out that
$\mathcal{G}_{\mathcal{R}_{+}}\simeq
\mathfrak{g}$,
$\mathcal{G}_{\mathcal{R}_{-}}\simeq
\mathfrak{g}_R$ where the algebra
$\mathfrak{g}_R$ is a linear space
$\mathfrak{g}$ equipped with the $R$-bracket \cite{RST2}
and $\mathcal{G}_{\mathcal{R}_{\pm}}\equiv \mathrm{Im}
\mathcal{R}_{\pm}$.


That is why our first observation is that using the standard
$R$-matrix scheme \cite{ST} applied to the Lie algebra
$\mathcal{G}$ equipped with the $R$-operator
$\mathcal{R}$ it is possible to obtain a set of commuting flows
on  extensions of $\mathfrak{g}$ by some
  Lie algebra $\mathfrak{a}$,  where
$\mathfrak{a}=\mathfrak{g}_R/J_R$ and
$J_R$ is an ideal in $\mathfrak{g}_R$. In the
particular case of
$J_R=\mathfrak{g}_R$  we rederive in a simple way the result of
\cite{SkrTMP2} about commutativity of the restrictions of the Casimir
 functions of $\mathfrak{g}$ onto the subalgebras
 $\mathfrak{g}_{R_{\pm}}$. In such a way we show that the results of \cite{SkrTMP2} fit  into the general $R$-matrix scheme. In the case $J_R=[\mathfrak{g}_R, \mathfrak{g}_R]$
we obtain an important generalization of the above mentioned
result, namely, we prove commutativity of the restrictions of the Casimir
 functions of $\mathfrak{g}$ onto the subalgebras
 $\mathfrak{g}_{R_{\pm}}$ shifted with the help of
 constant elements $c_{\mp} \in
 [\mathfrak{g}_{R_{\mp}}, \mathfrak{g}_{R_{\mp}}]$
 respectively. Let us note that, in this case, the obtained
 functions commute with respect to the Lie--Poisson bracket on
 $\mathfrak{g}$ ``shifted" by the constant
 element $c_{+}-c_{-}$.

A consideration of commutative families on more
complicated quotients (with non-abelian $\mathfrak{a}$) might
also be useful  in the theory of soliton equations. Indeed,
our second simple observation suggests that, whatever
quotient of $\mathcal{G}_{\mathcal{R}}$ one considers,
one may choose $M$-operators from the Lax equations
$$
\dot L=[L,M]
$$
corresponding to the commuting hamiltonians (the Casimir functions
restricted onto this quotient) to take the values in
$\mathfrak{g}$. From this it follows that one can
obtain zero-curvature equations with
$\mathfrak{g}$-valued $U$-$V$-pairs as a consistency
condition of the Lax equations  on
$\mathfrak{g}\ominus\mathfrak{a}$.

We illustrate the above method by the example of abelian
 \cite{Mikh1}, \cite{Mikh2}  and non-abelian  (see \cite{Raz} and references therein) Toda field equations, that
are naturally obtained in the framework of the above scheme if
$\mathfrak{g}$ is a loop algebra equipped with various
gradings. The corresponding quotient algebra
 in this case is the simplest
non-abelian extension of $\mathfrak{g}$ obtained in
the framework of the above construction. In the case if
$\mathfrak{g}=gl((\infty))$ equipped  with the natural
decomposition into a sum of two subalgebras, coinciding with upper
triangular and strictly lower triangular matrices, we recover
results of \cite{GCar} about Lie--Poisson structure and
Lie-theoretical interpretation of infinite-component Toda field
equations, its $U$-$V$ pair, auxiliary Lax pairs etc. \cite{Ueno}.

At the end of the paper for the sake of completeness we also
consider the prolongation of the second and third order Poisson
structures, existing for the cases of certain $R$-operators
$\mathfrak{g}$ on the classical double. It occurred
that the quadratic and cubic structures are always prolonged on
$\mathcal{G}$, whenever they exist on
$\mathfrak{g}$. Nevertheless  their usage in the
soliton theory is restricted by the fact that the described above
quotient spaces -- Poisson spaces of the linear
$\mathcal{R}$-bracket on $\mathcal{G}$ are not in
general Poisson subspaces of the quadratic and cubic brackets.

The structure of the present paper is the following: in the second
section we introduce main definition and notations. In the third
section we use classical double in order to obtain commutative
families  on $\mathfrak{g}^*$ and its extensions.  In
the fourth section we utilize the obtained results in order to
construct zero-curvature equations with
$\mathfrak{g}$-valued $U$-$V$ pairs and illustrate this
approach on the examples of abelian and non-abelian Toda field
equations. Finally in the fifth section we consider the
prolongation of the second and third order Poisson structures on
the double.

\section{Definitions and notations}
\subsection{Lie algebras and classical $R$-operators} Let
$\mathfrak{g}$ be a Lie algebra (finite or
infinite-dimensional) with a Lie bracket $[\ ,\ ]$,
$R:\mathfrak{g}\rightarrow \mathfrak{g}$
be a linear operator. The operator $R$ is called a classical
$R$-operator if it satisfies the modified Yang-Baxter equation:
\begin{equation*}
R([R(X),Y]+ [X,R(Y)])-[R(X),R(Y)]=[X,Y],\quad \forall X,Y \in
\mathfrak{g}.
\end{equation*}
Using classical $R$-operator it is possible to define another
bracket on $\mathfrak{g}$ by the formula \cite{ST}:
\begin{equation}\label{rlbd}
[X,Y]_R=[R(X),Y]+ [X,R(Y)],\quad X,Y \in \mathfrak{g}.
\end{equation}
We will denote $\mathfrak{g}_R$ the linear space $\mathfrak{g}$ equipped
with the Lie bracket $[\ ,\ ]_R$.

We will also use hereafter the following notations: $R_{\pm}\equiv
R\pm {\rm Id}.$

It is known \cite{ST} that the images $\mathfrak{g}_{R_{\pm}}=\mathrm{Im} R_{\pm}$ of the maps  $R_{\pm}$ define Lie subalgebras
$\mathfrak{g}_{R_{\pm}}\subset
\mathfrak{g}$. As it is
easy to see from their definition
$\mathfrak{g}_{R_{+}}+\mathfrak{g}_{R_{-}}=\mathfrak{g}$,
but, in general, this sum is  not a direct sum of vector spaces,
i.e., in the general case,
$\mathfrak{g}_{R_{+}}\cap\mathfrak{g}_{R_{-}}\neq
0$.

{\it Remark 1.} The situation is much simpler in the  case of
a Lie algebra $\mathfrak{g}$ with the so-called
``Adler--Kostant--Symes" (AKS) decomposition into a direct sum of two
Lie subalgebras:
$\mathfrak{g}={\mathfrak{g}_+} +
{\mathfrak{g}_-}$. Indeed, if $P_{\pm}$ are the projection
operators onto the subalgebras ${\mathfrak{g}_{\pm}}$
then \cite{ST} $R=P_+ - P_-$ is a classical $R$-matrix.
 It is easy to see that in this case $R_+={\rm Id}+R=2P_+$,
$R_-=R-{\rm Id
}=-2P_-$, are proportional to the projection operators onto
the  subalgebras $\mathfrak{g}_{\pm}$. It also follows
that $\mathfrak{g}_{R_{\pm}}\equiv
\mathfrak{g}_{\pm}$ and
$\mathfrak{g}_{R_{+}} \cap
\mathfrak{g}_{R_{-}}=0$. It is also known \cite{ST}
that in this case $$\mathfrak{g}_R=
\mathfrak{g}_{+}\ominus
\mathfrak{g}_{-}.$$

\subsection{Classical double} Let us now consider the  ``double"
of the Lie algebra $\mathfrak{g}$, namely the direct
sum algebra
$\mathcal{G}=\mathfrak{g}\oplus\mathfrak{g}.$
Let us identify the elements of
$\mathcal{X}\in\mathcal{G}$ with
vector columns $\mathcal{X}=\left(%
\begin{array}{c}
  X_1 \\
  X_2\\
\end{array}%
\right),$ where $X_i \in \mathfrak{g}$. The
bracket of two elements $\mathcal{X},\mathcal{Y}
\in \mathcal{G}$ is given by the standard formula $[\mathcal{X},
\mathcal{Y}]=
\begin{pmatrix}
  [X_1,Y_1] \\
  [X_2,Y_2]\\
\end{pmatrix}.$
The following construction has been developed in \cite{RST2}.

\begin{teo}
(i) Given an arbitrary classical $R$-operator on $\mathfrak{g}$, the operator defined on the double  by the formula
$$\mathcal{R}=
\begin{pmatrix}
  R & -R_- \\
  R_+ & -R \\
\end{pmatrix},$$
will be a classical $R$-operator on $\mathcal{G}$. (ii) The corresponding $R$-bracket $[\ ,\
]_{\mathcal{R}}$ on $\mathcal{G}$ has the form:
\begin{equation}
[\mathcal{X}, \mathcal{Y}]_{\mathcal{R}}=
\begin{pmatrix}
  [X_1,Y_1]_R -([X_1,R_-(Y_2)] + [R_-(X_2),Y_1])\\
  -[X_2,Y_2]_R+ ([X_2,R_+(Y_1)] + [R_+(X_1),Y_2])\\
\end{pmatrix}.
\end{equation}
(iii) The $R$-matrix
$\mathcal{R}$ is of the Adler--Kostant--Symes  type.
\end{teo}

For convenience of the reader we give a sketch of the proof of the
third item of the Theorem. Let us consider the following
operators:
$$\mathcal{R}_+=\mathcal{R}+{\rm id}=
\begin{pmatrix}
  R_+ & -R_- \\
  R_+ & -R_- \\
\end{pmatrix},\   \mathcal{R}_-=\mathcal{R}-{\rm id}=
\begin{pmatrix}
  R_- & -R_- \\
  R_+ & -R_+ \\
\end{pmatrix}.$$
Denote $\mathcal{G}_{\mathcal{R}_{\pm}}=\mathrm{Im} R_{\pm}$ the corresponding Lie subalgebras. It is
easy to see that $$\mathcal{R}_+(\mathcal{X})=
\begin{pmatrix}
  R_+(X_1)-R_-(X_2)\\
  R_+(X_1)-R_-(X_2)\\
\end{pmatrix},\ \mathcal{R}_-(\mathcal{X})=
\begin{pmatrix}
  R_-(X_1-X_2)\\
  R_+(X_1-X_2)\\
\end{pmatrix}.$$
From this it follows that
$\mathcal{G}_{\mathcal{R}_{+}}\equiv\mathcal{G}_d
\simeq \mathfrak{g}$,
$\mathcal{G}_{\mathcal{R}_{-}}\simeq
\mathfrak{g}_R$ where
$$\mathcal{G}_d=\left\{\left(%
\begin{array}{c}
  X \\
  X \\
\end{array}%
\right)|  X\in \mathfrak{g}\right\},\
\mathcal{G}_{\mathcal{R}_{-}}=\left\{\left(%
\begin{array}{c}
  R_-(X) \\
  R_+(X) \\
\end{array}%
\right)| X\in \mathfrak{g} \right\}.$$
 It is easy to see that
$\mathrm{Ker} \mathcal{R}_+= \mathrm{Im} \mathcal{R}_-$ and
$\mathrm{Ker} \mathcal{R}_-= \mathrm{Im} \mathcal{R}_+$. Hence the
decomposition $\mathcal{G}=
\mathcal{G}_{\mathcal{R}_{+}}+
\mathcal{G}_{\mathcal{R}_{-}}$ is a decomposition into
a direct sum of vector spaces and operator $\mathcal{R}$ is of the
AKS type, i.e.
$\mathcal{R}=(\mathcal{P}_{\mathcal{G}_{\mathcal{R}_{+}}}-
\mathcal{P}_{\mathcal{G}_{{R}_-}})$. This fact  also
follows from the  easily proved identities
$\mathcal{R}^2_{\pm}=2\mathcal{R}_{\pm},\
\mathcal{R}_{+}\mathcal{R}_{-}=0$ implying that
$\mathcal{R}_{\pm}$ are proportional to projection operators and
their  images  do not intersect. That is why we have that
$$\mathcal{G}_{\mathcal{R}}=\mathcal{G}_{\mathcal{R}_{+}}\ominus
\mathcal{G}_{{R}_-}.$$ For the ${\mathcal R}$-bracket on the double this identity means that $$[\mathcal{X},
\mathcal{Y}]_{\mathcal{R}}=2([\mathcal{X}_+, \mathcal{Y}_+]-
[\mathcal{X}_-, \mathcal{Y}_-]),$$ where $\mathcal{X}_+\equiv
\mathcal{R}_+(\mathcal{X}),$ $\mathcal{X}_-\equiv
\mathcal{R}_-(\mathcal{X})$ etc. or, more explicitly
\begin{equation*}
[\mathcal{X}, \mathcal{Y}]_{\mathcal{R}}=
\begin{pmatrix}
  [(R_+(X_1)-R_-(X_2)),(R_+(Y_1)-R_-(Y_2))]\\
  [(R_+(X_1)-R_-(X_2)),(R_+(Y_1)-R_-(Y_2))]\\
\end{pmatrix}
-
\begin{pmatrix}
  [(R_-(X_1-X_2)),R_-(Y_1-Y_2)]\\
  [(R_+(X_1-X_2)),R_+(Y_1-Y_2)]\\
\end{pmatrix}.
\end{equation*}

{\it Remark 2.} In the case of the Adler--Kostant--Symes $R$-operators
all the formulas of this subsection are substantially simplified.
In particular, the action of the $R$-operator $\mathcal{R}$ on the
element $\mathcal{X}$
is given by the formula: $\mathcal{R}(\mathcal{X})=\left(%
\begin{array}{c}
 X^+_1-X^-_1+2X^-_2 \\
  2X^+_1-X^+_2+X^-_2\\
\end{array}%
\right)$, the $R$-bracket (\ref{rlbd}) is written as follows:
\begin{equation*}
[\mathcal{X}, \mathcal{Y}]_{\mathcal{R}}=
\begin{pmatrix}
  [X^+_1,Y^+_1]- [X^-_1,Y^-_1] -([X_1,Y^-_2] + [X^-_2,Y_1])\\
  -[X^+_2,Y^+_2]+[X^-_2,Y^-_2] + ([X_2,Y^+_1] + [X^+_1,Y_2])\\
\end{pmatrix},
\end{equation*}
where $X_i=X_i^++X_i^-$, $Y_i=Y_i^++Y_i^-$ and
$X_i^{\pm}=P_{\pm}(X)$, $Y_i^{\pm}=P_{\pm}(Y)$.

\subsection{Dual spaces, Lie--Poisson
brackets and invariant functions} Let $\mathfrak{g}^*$
be the dual space to $\mathfrak{g}$ and
 $\langle \ , \ \rangle:
\mathfrak{g}^*\times
\mathfrak{g}\rightarrow \mathbb{C}$ be the natural pairing
between $\mathfrak{g}^*$ and
$\mathfrak{g}$. Let $\{X_i\, | i\in I \}$ be a basis in
the Lie algebra $\mathfrak{g}$, where the set $I$ is
finite for the case of finite dimensional Lie algebra and
countable in the infinite-dimensional case. Let $\{X^*_i| i\in
I\}$, $\langle X^*_j, X_i \rangle =\d_{ij}$, be a basis in the
dual space $\mathfrak{g}^*$. Let $L=\sum\limits_{i}
L_iX^*_i \in \mathfrak{g}^*$ be a generic element of
$\mathfrak{g}^*$, $L_i$ be the coordinate functions on
$\mathfrak{g}^*$.  Let us consider the standard Lie--Poisson
bracket between $F_1, F_2 \in
C^{\infty}(\mathfrak{g}^*)$:
\begin{equation*}
\{F_1(L),F_2(L)\}= \langle L ,[\nabla F_1, \nabla F_2] \rangle,
\end{equation*}
where  $\nabla F_k(L)= \sum\limits_{i \in I} \dfrac{\partial
F_k(L)}{\partial L_i}X_i$ is a so-called algebra-valued gradient
of $F_k$. Like in \cite{ST} an $R$-operator provides us with the
 so-called ``R-bracket" on $\mathfrak{g}^*$ in the following
way
\begin{equation}\label{rpb}
\{F_1(L),F_2(L)\}_R= \langle L ,[\nabla F_1, \nabla F_2]_R
\rangle,
\end{equation}

Let us consider the dual space $\mathcal{G}^*$ to the double. We
identify its elements $\mathcal{L}\in \mathfrak{g}^*\oplus \mathfrak{g}^*$
with the vector columns: $\mathcal{L}=
\left(%
\begin{array}{c}
  L_1 \\
  L_2 \\
\end{array}%
\right)$
 where $L_1, ~ L_2 \in \mathfrak{g}^*$ and the pairing between $\mathcal{L}\in\mathcal{G}^*$ and $\mathcal{X}\in \mathcal{G}$ is defined as follows:
$$\langle \mathcal{L}, \mathcal{X}\rangle =
  \langle L_1,X_1\rangle +\langle L_2,X_2\rangle.$$
It is also defined the original Lie--Poisson bracket
on $\mathcal{G}^*$
\begin{equation}\label{pbd}
\{F_1(\mathcal{L}),F_2(\mathcal{L})\}= \langle \mathcal{L}
,[\widetilde{\nabla} F_1, \widetilde{\nabla} F_2] \rangle,
\end{equation}
and $R$-bracket on $\mathcal{G}^*$ corresponding to
the $R$-operator $\mathcal{R}$
\begin{equation}\label{rpbd}
\{F_1(\mathcal{L}),F_2(\mathcal{L})\}_{\mathcal{R}}= \langle
\mathcal{L} ,[\widetilde{\nabla} F_1, \widetilde{\nabla}
F_2]_{\mathcal{R}} \rangle
\end{equation}
where $$\widetilde{\nabla} F= \begin{pmatrix}
\nabla_1 F\\
 \nabla_2 F\\
\end{pmatrix}$$  and $\nabla_{1, 2} F$ is the algebra-valued gradient
with respect to  the variable $L_{1, 2}$.

Denote $R^*$ the adjoint operator to $R$,
$$
R^*: \mathfrak{g}^*\to \mathfrak{g}^*, \quad \langle R^{*}(L),X\rangle\equiv
\langle L,R(X)\rangle.
$$
It is easy to see that the adjoint
operators  to $\mathcal{R}_{\pm}$  have the following form:
\begin{equation}\label{rplus}
 \mathcal{R}^*_+(\mathcal{L})=
 \left(%
\begin{array}{c}
 R_+^*(L_1)+R^*_+(L_2) \\
  -(R^*_-(L_1)+R^*_-(L_2)) \\
\end{array}%
\right), \ \
   \mathcal{R}^*_-(\mathcal{L})=
\left(%
\begin{array}{c}
  R^*_-(L_1)+R^*_+(L_2) \\
 -(R^*_-(L_1)+R^*_+(L_2))\\
\end{array}%
\right)
  \end{equation}
We will need these explicit formulas below while constructing
Poisson-commuting integrals.

In the subsequent considerations we will also need to know the explicit form of
Casimir functions of $\mathcal{G}$. In more details,
let $I(L)\in I^{G}(\mathfrak{g}^*)$ be a Casimir
function of $\mathfrak{g}$, i.e. $$ \{I(L), F(L)
\}=0\quad \forall F(L)\in S(\mathfrak{g}^*).$$
Denote $\{ I_k(L)\}_{k\in K}$ the set of generators of the ring of Casimirs of $\mathfrak{g}^*$. Here the set of labels $K$ is infinite if the Lie algebra is infinite-dimensional.

\begin{lem} The ring of Casimir functions of $\mathcal{G}$ is generated by
the following functions
\begin{equation}\label{dcas}
I_{k,1}(\mathcal{L})\equiv I_k(L_1),\ I_{k,2}(\mathcal{L})\equiv
I_k(L_2), \quad k\in K.
\end{equation}
\end{lem}

The proof is straightforward.

Let us say few words about quadratic Casimirs and commuting functions that
are obtained with their help. Let $(\ ,\ )$ be an invariant form
on $\mathfrak{g}$. With the help of the latter we can identify
$\mathfrak{g}$ and  $\mathfrak{g}^*$.
Then we have the following obvious second-order Casimir functions
(or generating functions of formal Casimirs in the case of loop
algebras)
$$I_{2,1}= \frac{1}{2}(L_1,L_1), \  I_{2,2}= \frac{1}{2}(L_2,L_2).$$

\section{Classical  double and commuting flows.}
In order to pass to the main construction of the present
article let us remind the
following theorem that can be obtained from the general theory of $R$-brackets
\cite{ST}
 applied to the case of the classical double.
\begin{teo}\label{rmi}
(i) The Casimir functions $I_{k,\e}(\mathcal{L})$ of the Lie--Poisson brackets of
$\mathcal{G}$ commute with respect to the brackets
$\{\ ,\  \}_{\mathcal{R}}$ on $\mathcal{G}^*$.

\noindent (ii)
The hamiltonian flows
$$
\frac{d}{dt^\epsilon_k} F({\mathcal L})=\left\{ F({\mathcal L}), I_{k,\epsilon}\right\}_{\mathcal R}, \quad k\in K, \quad \epsilon=1, 2
$$
generated by the Casimir invariants
$I_{k,\e}$  are written in the Lax-type form:
\begin{equation}\label{Lax}
\frac{d \mathcal{L}}{d t^{\e}_k}= ad^*_{\mathcal{R}_{+}
\widetilde{\nabla} I_{k,\e} } \mathcal{L}.
\end{equation}
\end{teo}

 Now, let us
briefly consider  commuting flows  on $\mathfrak{g}$
and  its extensions that can be obtained using the theory of
classical double. For this purpose let us remind that, using the
fact that the projection onto the quotient algebra is a canonical
homomorphism, one can deduce the  following
\begin{nas}
Let $J$ be an ideal in $\mathcal{G}_{\mathcal{R}}$.
Denote $\pi: \mathcal{G} \rightarrow{ \mathcal{G}}/
J $ the projection onto the quotient algebra. Let
$\pi^*: (\mathcal{G}/ J )^*\rightarrow
\mathcal{G}^*
 $ be  the dual map. Then:

(i) The functions $I_{k,\e}(\pi^*(\mathcal{L}))$ of
$\mathcal{G}$ commute with respect to the brackets
$\{\ ,\  \}_{\mathcal{R}}$ on
$(\mathcal{G}/J)^*$.

(ii)  The hamiltonian flows corresponding to functions
$I_{k,\e}(\pi^*(\mathcal{L}))$  are written in the Lax-type form:
\begin{eqnarray}\label{pLax}
&&
\pi^*(\frac{d \mathcal{L}}{d t^{\e}_k})= ad^*_{M_{k,\e}}
\pi^*(\mathcal{L}), \quad k\in K, \quad \e=1, \, 2
\\
&&
M_{k,\e}=\mathcal{R}_{+}(\pi
\widetilde{\nabla} I_{k,\e}(\pi^*(\mathcal{L})).
\nonumber
\end{eqnarray}
\end{nas}

Let us now assume that there exist  non-trivial ideals
$J_{R_{\pm}}\subset
\mathfrak{g}_{R_{\pm}}$ such that the quotients
$\mathfrak{g}_{R_{\pm}}/J_{R_{\pm}}$ are
finite-dimensional. In such a case, applying the above Corollary  \ref{rmi} and taking the quotient over the
ideal $J=J_{R_{+}}+J_{R_{-}}$
we will obtain a Poisson-commuting set of functions on the dual space to the finite-dimensional
extensions of $\mathfrak{g}$. Indeed, we have
$$\mathcal{G}_{\mathcal{R}}/J=
(\mathcal{G}_{\mathcal{R}_+} \ominus
\mathcal{G}_{\mathcal{R}_-})/J \simeq
\mathfrak{g} \ominus \mathfrak{a}, \text{ where }
\mathfrak{a}\simeq
\mathcal{G}_{\mathcal{R}_-}/J.$$

{\it Remark 3.} For the quotient algebras described above the
$M$-operators from the Lax equations (\ref{pLax}) have the
following form:
$$M_{k,1}=\left(%
\begin{array}{c}
{R}_{+}({\nabla}
I_{k,1}(\pi^*(\mathcal{L})) \\
{R}_{+}({\nabla}
I_{k,1}(\pi^*(\mathcal{L}))\\
\end{array}%
\right), \ \  M_{k,2}=-\left(%
\begin{array}{c}
{R}_{-}({\nabla}
I_{k,2}(\pi^*(\mathcal{L}))\\
{R}_{-}({\nabla}
I_{k,2}(\pi^*(\mathcal{L}))\\
\end{array}%
\right),$$
 i.e. they belong to the diagonal subalgebra and may be
identified with the elements of $\mathfrak{g}$.

This fact will be used when constructing zero-curvature equations
with values in $\mathfrak{g}$. Note that
 the corresponding Lax equations (\ref{pLax}),
dynamical variables, Poisson brackets etc. are written on
the double of $\mathfrak{g}$. Note also that the
absence of the operator $\pi$ in the above formulas for the
$M$-operators is explained by the fact that the ideal
$J$ in this case was chosen in such a way that
$\mathcal{R}_{+}\pi=\mathcal{R}_{+}$.

{\it Example 1.} Let us consider the case of Adler--Kostant--Symes
$R$-operators: $R=P_+-P_-$. Let us describe more explicitly the
quotient algebras
$\mathcal{G}_{\mathcal{R}}/J$ and the
corresponding dual spaces. We have $\mathfrak{g}=
\mathfrak{g}_++ \mathfrak{g}_-$,
$\mathfrak{g}_{R_{\pm}}=
\mathfrak{g}_{\pm}$ and $J_{R_{\pm}}\equiv
J_{\pm}$ are ideals in
$\mathfrak{g}_{\pm}$. Elements of the corresponding
quotient $\mathcal{G}_{\mathcal{R}}/J$,
where and $J=J_+ + J_-$, have
the form: $\mathcal{X}=
\left(%
\begin{array}{c}
  X_1^+ + X_1^{-'} \\
 X_2^- + X_2^{+'} \\
\end{array}%
\right),$ where $X_1^{-'} \in
\mathfrak{g}_-/J_-$, $X_2^{+'} \in
\mathfrak{g}_+/J_+$.
 The corresponding dual space consists of the elements
$\mathcal{L}= \left(%
\begin{array}{c}
 L_1^+ + L_1^{-'} \\
 L_2^-+L_2^{+'} \\
\end{array}%
\right)$, where $L_1^{-'} \in
(\mathfrak{g}_-/J_-)^*$, $L_1^{+} \in
(\mathfrak{g}_+)^*$, $L_2^{-} \in
(\mathfrak{g}_-)^*$, $L_2^{+'} \in
(\mathfrak{g}_+/J_+)^*$.

In the next subsections  we will consider in more details  the
cases of $\mathfrak{a}\simeq
\mathcal{G}_{\mathcal{R}_-}/J=0$ and of an abelian $\mathfrak{a}$ that lead
to the commutative algebras of integrals on
$\mathfrak{g}^*$ itself.

\subsection{``Dual" $R$-matrix commutativity.}
 Let us at first consider the  consequences of
the general Theorem \ref{rmi}  to a construction of
Poisson-commuting sets on $\mathfrak{g}^*$ with
respect to the standard Lie--Poisson brackets $\{\ ,\ \}$.
 The
following Theorem holds true:
\begin{teo}\label{drmi}
(i) The functions $I_k(R^*_{\pm}(L))$ on
$\mathfrak{g}^*$ generate an abelian subalgebra in
$C^{\infty}(\mathfrak{g}^*)$ with respect to the
Lie--Poisson
brackets $\{\ , \}$ on $\mathfrak{g}^*$:\\

 $\{I_k(R^*_{+}(L)),I_l(R^*_{+}(L))\}=0$,
$\{I_k(R^*_{-}(L)),I_l(R^*_{-}(L))\}=0$,
$\{I_k(R^*_{+}L),I_l(R^*_{-}(L))\}=0$.\\

(ii) The hamiltonian  equations corresponding to the hamiltonians
$I^{R_{\pm}}_k(L)$ are written in the Lax-type form:
\begin{eqnarray}\label{Lax2}
&&
\frac{d L}{d t^{\pm}_k}= ad^*_{M_k^\pm } L
\\
&&
M_k^\pm =\nabla I_k(R^*_{\pm}(L)).
\nonumber
\end{eqnarray}
\end{teo}
{\it Proof.} In order to prove the item $(i)$ of the theorem  let
us project the functions $I_{k,\e}(\mathcal{L})$, $\e\in
\overline{1,2}$
 onto the dual space to the quotient algebra $\mathcal{G}_{\mathcal{R}}/
 \mathcal{G}_{R_-}$ isomorphic to the subalgebra  $\mathcal{G}_{R_+}$. Due to the explicit formulas (\ref{rplus}),
 (\ref{dcas}) we obtain the following expressions for the projected Casimirs:
$$I_{k,1}(P^*_{\mathcal{G}_{R_+}}(\mathcal{L)})=I_k(R_+^*(L_1+L_2)),\
I_{k,2}(P^*_{\mathcal{G}_{R_+}}(\mathcal{L)})=I_k(R_-^*(L_1+L_2)),$$
where we have  used that
$P^*_{\mathcal{G}_{R_+}}=\frac{1}{2}\mathcal{R}^*_+$
and assumed that $I_k$ are homogeneous function of $L$.

Let us observe that, due to the
fact that $\mathcal{R}_{+}$ is projection operator
 and the corresponding $R$-operator $\mathcal{R}$ is of the Adler--Kostant--Symes type we have:
$$\{F(\mathcal{R}^*_{+}(\mathcal{L})),G(\mathcal{R}^*_{+}(\mathcal{L}))\}=
\{F(\mathcal{R}^*_{+}(\mathcal{L})),G(\mathcal{R}^*_{+}(\mathcal{L}))\}_{\mathcal{R}}.$$
Using the fact that the projection onto the quotient algebra is a
canonical homomorphism we  obtain:
$$\{F(\mathcal{R}^*_{\pm}(\mathcal{L})),G(\mathcal{R}^*_{\pm}(\mathcal{L}))\}_{\mathcal{R}}=
\left(\{F(\mathcal{L}),G(\mathcal{L})\}_{\mathcal{R}}\right)|_{\mathcal{L}=\mathcal{R}^*_{\pm}(\mathcal{L})}.$$
Hence, putting $F=I_{k,\e}$, $G=I_{l,\e'}$, we will have:
$$\{I_{k,\e}(\mathcal{R}^*_+(\mathcal{L)}),
I_{l,\e'}(\mathcal{R}^*_+(\mathcal{L)})\}=(\{I_{k,\e}(\mathcal{L}),
I_{l,\e'}(\mathcal{L})\}_{\mathcal{R}})|_{\mathcal{L}=\mathcal{R}^*_{+}(\mathcal{L})},$$
where $\e,\e' \in \overline{1,2}$. On the other hand,
$\{I_{k,\e}(\mathcal{L}),
I_{l,\e'}(\mathcal{L})\}_{\mathcal{R}}=0$ by virtue of the Theorem
\ref{rmi}.

Taking into account the explicit form of the functions
$I_{k,\e}(\mathcal{L})$ we finally obtain
$$\{ I_k(R_+^*(L_1+L_2)), I_l(R_+^*(L_1+L_2))\}=0, \{I_k(R_+^*(L_1+L_2)), I_l(R_-^*(L_1+L_2))\}=0,$$
$$\{ I_k(R_-^*(L_1+L_2)), I_l(R_-^*(L_1+L_2))\}=0.$$
Now, in order to prove the item $(i)$ of the theorem it remains to
observe that elements of the form $L\equiv L_1+L_2$ belong to the subspace
$(\mathcal{G}_d)^*$ and the corresponding coordinate
functions constitute Lie algebra isomorphic to
$(\mathfrak{g},\{\ ,\ \})$ with respect to the initial
Lie--Poisson brackets on $\mathcal{G}$.

Item $(ii)$ of the Theorem can be proven using the part $(ii)$ of the
Theorem  \ref{rmi}. It can also be proven by noticing that any
hamiltonian equation on $\mathfrak{g}^*$ is re-written
in the Euler--Arnold form.

Theorem is proven.

{\it Remark 4.} In the paper \cite{SkrTMP1} the above theorem was
proven directly without any appeal to the classical double.
Nevertheless the proof using the classical double is more simple
and makes the Theorem \ref{drmi} fit into the general $R$-matrix
scheme.

{\it Example 2.} Let us now consider in more details the case
$R=P_+-P_-$ and more explicitly describe Lie algebra
$\mathcal{G}_{\mathcal{R}_+}$ realized as a quotient
algebra. We have  $\mathrm{Ker} \mathcal{R}_+= \mathrm{Im}
\mathcal{R}_-$  and $\mathrm{Im} \mathcal{R}_-=
\left(%
\begin{array}{c}
  X_1^- \\
 X_2^+  \\
\end{array}%
\right). $ The corresponding quotient algebra
$\mathcal{G}_{\mathcal{R}}/\mathrm{Im}\mathcal{R}_-$
can be identified with the linear space consisting of the
following elements
$$\mathcal{X}=
\left(%
\begin{array}{c}
  X_1^+ \\
 X_2^-  \\
\end{array}%
\right).
$$
Such the space is
 isomorphic to $\mathcal{G}_{\mathcal{R}_+}$
and the isomorphism is established by
the map $\mathcal{R}_+$: $$\left(%
\begin{array}{c}
  X_1^+ \\
 X_2^-  \\
\end{array}%
\right) \rightarrow
\left(%
\begin{array}{c}
  X_1^+-X_2^- \\
 X_1^+-X_2^-  \\
\end{array}%
\right)$$

 The corresponding dual space consists of the elements
$\mathcal{L}= \left(%
\begin{array}{c}
  L_1^+ \\
 L_2^-  \\
\end{array}%
\right)$. Observe that such elements can be identified with
the elements $L=L_1^++L_2^-$ of the linear space $\mathfrak{g}^*$. Moreover, the corresponding Lie--Poisson brackets of these
elements on
$\mathcal{G}_{\mathcal{R}}/\mathrm{Im}\mathcal{R}_-$
coincide with the standard Lie--Poisson bracket  of the element
$L=L^++L^-$ on $\mathfrak{g}^*$.

The Casimir functions $I_{k,1}(\mathcal{L})$ and $I_{k,2}(\mathcal{L})$ restricted to the dual space of the quotient
algebra are simply the functions $I_k(L_1^+)$ and $I_k(L_2^-)$. After
the above identification they reduce to the functions $I_k(L^+)$ and
$I_k(L^-)$ on $\mathfrak{g}^*$.

\subsection{``Shift of the argument"  and  commutative algebras.}
Let us now consider   commutative subalgebras of functions on
$\mathfrak{g}^*$ generalizing the commutative algebras
constructed in the previous subsection that depend on
additional parameters and can be obtained using the theory of
classical double.

The method that  allows one to introduce additional
parameters into commutative subalgebras is the so-called shift of the
argument. The following theorem holds true.

\begin{teo}\label{sdrmi}
Let $c_{\pm}$ be  constant elements of
$\mathfrak{g}_{R_{\pm}}^*$ such that $c_{\pm} \bot
([\mathfrak{g}_{R_+},
\mathfrak{g}_{R_+}]\bigcup
[\mathfrak{g}_{R_-}, \mathfrak{g}_{R_-}])$
and $I_k(L)$, $I_l(L)$ be Casimir functions of
$\mathfrak{g}$. Then (i)
\begin{gather}
\{I_k(R^*_{+}(L)+c_-),I_l(R^*_{+}(L)+c_-)\}_{c} =0,
\{I_k(R^*_{-}(L)+c_+),I_l(R^*_{-}(L)+c_+)\}_{c}=0,\\
\{I_k(R^*_{+}(L)+c_-),I_l(R^*_{-}(L)+c_+)\}_{c}=0,
\end{gather}
where $\{\ ,\ \}_{c}$ is the shifted bracket:
\begin{equation}\label{clpb}
\{F_1(L),F_2(L)\}_{c}= \langle L ,[\nabla F_1, \nabla F_2]
\rangle+\langle c_--c_+ ,[\nabla F_1, \nabla F_2] \rangle
\end{equation}
(ii) The corresponding hamiltonian equations are written in the
Euler-Arnold form:
\begin{equation}\label{Lax4}
\frac{d L}{d t^{\pm}_k}= ad^*_{\nabla I_k(R^*_{\pm}(L)+c_{\mp })}
(L+c_--c_+).
\end{equation}
\end{teo}

{\it Proof.} In order to prove item $(i)$ of this theorem let us
take into account that $\mathcal{G}_{\mathcal{R}}=
\mathcal{G}_{\mathcal{R}_+}\ominus
\mathcal{G}_{\mathcal{R}_-}$. Hence,
$[\mathcal{G}_{\mathcal{R}},
\mathcal{G}_{\mathcal{R}}]=[\mathcal{G}_{\mathcal{R}_+},\mathcal{G}_{\mathcal{R}_+}]
\ominus [\mathcal{G}_{\mathcal{R}_-},
\mathcal{G}_{\mathcal{R}_-}]$. Let us explicitly
describe the ideal $[\mathcal{G}_{\mathcal{R}_-},
\mathcal{G}_{\mathcal{R}_-}]$. We have
$\mathcal{G}_{\mathcal{R}_-}=\{\left(%
\begin{array}{c}
  X_1\\
 X_2 \\
\end{array}\right)
|\text{ where } X_1 \in \mathfrak{g}_{R_-}, X_2 \in
\mathfrak{g}_{R_+} \}$. From this it follows that
element
 $C=(c_1,c_2) \in \mathcal{G}^{*}$ is orthogonal  to
 $[\mathcal{G}_{\mathcal{R}_-},\mathcal{G}_{\mathcal{R}_-}]$ if
 $c_1 \bot [\mathfrak{g}_{R_-},\mathfrak{g}_{R_-}]$, $c_2 \bot
 [\mathfrak{g}_{R_+},\mathfrak{g}_{R_+}]$. On the other hand, as
 it follows  from the explicit form of the elements of
 $\mathcal{G}_{\mathcal{R}_-}^*$, $c_2=-c_1=-c$.
Hence $C=(c,-c)$ is an element of the dual space to the Lie subalgebra
$\mathcal{G}_{\mathcal{R}_-}$. It is orthogonal to
$[\mathcal{G}_{\mathcal{R}_-},\mathcal{G}_{\mathcal{R}_-}]$
if $c \bot ([\mathfrak{g}_{R_+},
\mathfrak{g}_{R_+}]\bigcup
[\mathfrak{g}_{R_-},
\mathfrak{g}_{R_-}])$.

That is why, factorizing the Lie algebra
$\mathcal{G}_{\mathcal{R}}$ over the ideal
$[\mathcal{G}_{\mathcal{R}_-},\mathcal{G}_{\mathcal{R}_-}]$
and taking into account that projection onto the quotient algebra
is a canonical homomorphism, applying the Theorem \ref{rmi} to
the Casimir functions $I_{k,\e}$,  $I_{l,\e'}$ we obtain that
\begin{gather*}
\{I_k(R^*_{+}(L_1+L_2)+c),I_l(R^*_{+}(L_1+L_2)+c)\}=
\{I_k(R^*_{-}(L_1+L_2)+c),I_l(R^*_{-}(L_1+L_2)+c)\}=0,\\
\{I_k(R^*_{+}(L_1+L_2)+c),I_l(R^*_{-}(L_1+L_2)+c)\}=0.
\end{gather*}
In such a way we have obtain a commutative subalgebra with a shift
element $c$ entering symmetrically in both ``positive" and
``negative" integrals, i.e. $c$ has components belonging both to
$\mathfrak{g}_{R_-}^*$ and
$\mathfrak{g}_{R_+}^*$. Let us also note, that shift
of the parts  of the Lax matrices belonging to
$\mathfrak{g}_{R_{\pm}}^*$ by the constant element of
the very same $\mathfrak{g}_{R_{\pm}}^*$ can be
eliminated by changing the variables. But this will lead to a
redefinition of the Poisson brackets. Making such a shift and
putting $c_{\pm}=\pm R^*_{\pm}(c)$, $L=L_1+L_2$ we obtain item
$(i)$ of the Proposition.

Item $(ii)$ is proved in an analogous way to the item $(ii)$ of the
previous Theorem.

Theorem is proven.

{\it Example 3.}  Let us, like in the previous Examples,  consider
the case of Adler-Kostant-Symes $R$-operators: $R=P_+-P_-$. Let
us more explicitly describe the quotient algebras
$\mathcal{G}_{\mathcal{R}}/[\mathcal{G}_{{\mathcal{R}}_-},\mathcal{G}_{{\mathcal{R}}_-}]$
and the corresponding dual spaces. We have
$[\mathcal{G}_{{\mathcal{R}}_-},\mathcal{G}_{{\mathcal{R}}_-}]=
[\mathfrak{g}_{+},
\mathfrak{g}_{+}]\ominus
[\mathfrak{g}_{-}, \mathfrak{g}_{-}]$.
Elements of the corresponding quotients
$\mathcal{G}_{\mathcal{R}}/[\mathcal{G}_{{\mathcal{R}}_-},\mathcal{G}_{{\mathcal{R}}_-}]$
 have the form: $\mathcal{X}=
\left(%
\begin{array}{c}
  X_1^+ + X_1^{-'} \\
 X_2^- + X_2^{+'} \\
\end{array}%
\right)$ where $X_1^{-'} \in
\mathfrak{g}_-/[\mathfrak{g}_{-},
\mathfrak{g}_{-}]$, $X_1^{+} \in
\mathfrak{g}_+$, $X_2^{-} \in
\mathfrak{g}_-$,  $X_2^{+'} \in
\mathfrak{g}_+/[\mathfrak{g}_{+},
\mathfrak{g}_{+}]$.
 The corresponding dual space consists of the elements
$\mathcal{L}= \left(%
\begin{array}{c}
 L_1^+ + L_1^{-'} \\
 L_2^-+L_2^{+'} \\
\end{array}%
\right)$, where $L_1^{-'} \in
(\mathfrak{g}_-/[\mathfrak{g}_{-},
\mathfrak{g}_{-}])^*$, $L_2^{+'} \in
(\mathfrak{g}_+/[\mathfrak{g}_{+},
\mathfrak{g}_{+}])^*$ . The elements $L_1^{-'}$,
$L_2^{+'}$ are constant with respect to the Poisson brackets on
$\mathcal{G}_{\mathcal{R}}/[\mathcal{G}_{{\mathcal{R}}_-},\mathcal{G}_{{\mathcal{R}}_-}]$
and we can put $$c_-=L_1^{-'},\quad  c_+=L_2^{+'}.$$

The Casimir functions restricted to the dual space of the quotient
algebra are functions $I_k(L_1^++ L_1^{-'})$ and
$I_l(L_2^-+L_2^{+'})$. Using the same  arguments as in the Example
2 and the making the same identification we can write them as
$I_k(L^++c_-)$ and $I_l(L^-+c_+)$ where $L$ is a generic element
of $\mathfrak{g}^*$. By virtue of the above
theorem they commute with respect to the brackets $\{\ ,\ \}_{c}$
on $\mathfrak{g}^*$:
\begin{gather*}
\{I_k(L^++c_-),I_l(L^++c_-)\}_{c} =0,\quad
\{I_k(L^-+c_+),I_l(L^-+c_+)\}_{c}=0,\\
\{I_k(L^++c_-),I_l(L^-+c_+)\}_{c}=0,
\end{gather*}
where the shifted bracket $\{\ ,\ \}_{c}$ is defined with the help
of the formula (\ref{clpb}).

\section{Integrable hierarchies and negative flows}
\subsection{Zero curvature equations}
 Let us remind one of the Lie algebraic
approaches to the theory of soliton equations \cite{New}. It is
based on the  zero-curvature conditions and its interpretation as
a consistency condition of two commuting Lax flows.

\begin{teo}\label{zceprop}
Let  ${\mathfrak{g}}$ be an infinite-dimensional Lie
algebra of ${\mathfrak{a}}$-valued meromorphic functions of one
complex variable and $\mathfrak{a}$ be a simple Lie algebra. Let
$H_i$ be Poisson-commuting polynomial functions on
$\mathfrak{g}^*$
$$\{H_1, H_2\}=0$$
where $\{\ ,\ \}$
is a standard Lie--Poisson brackets on
${\mathfrak{g}}^*$.
Then their ${\mathfrak{g}}$-valued gradients satisfy
 zero-curvature equation:
\begin{equation}\label{zce}
\frac{\partial \nabla H_1}{\partial t_2} - \frac{\partial \nabla
H_2}{\partial t_1} + [\nabla H_1, \nabla H_2]= 0
\end{equation}
and $t_i$ are parameters along the trajectories of hamiltonian
equations
$$
\frac{\partial}{\partial t_i} =\{ ~\cdot ~, H_i\}, \quad i=1, \, 2
$$
generated by the hamiltonians $H_i$.
\end{teo}
{\it  Proof.} In order to prove the Theorem let us first observe
that the hamiltonian equations on $\mathfrak{g}^*$
corresponding to the hamiltonians $H_i$ and the standard Lie--Poisson
brackets are always written in the Euler--Arnold (generalized Lax)
form:
$$\frac{\partial L}{\partial t_i}= ad_{\nabla H_i }^* L,$$
where $\nabla H_s$ is an algebra-valued gradient of $H_s$, i.e.
$\nabla H_s=\sum\limits_{i} \dfrac{\partial H_s}{\partial
L_i}X_i$, and $L= \sum\limits_{i} L_iX^*_i$ is  a generic element
of the dual space $\mathfrak{g}^*$.
 Using the commutativity of the time flows corresponding to
 $s=1,2$ one derives the following identity:
$$ad_{\left(\dfrac{\partial \nabla H_1}{\partial t_2} - \dfrac{\partial \nabla
H_2}{\partial t_1} + [\nabla H_1, \nabla H_2]\right)}^* L=0.$$

From this follows that
\begin{equation}\label{modzce}
\frac{\partial \nabla H_1}{\partial t_2} - \frac{\partial \nabla
H_2}{\partial t_1} + [\nabla H_1, \nabla H_2]= k \nabla I,
\end{equation}
where $I$ is a Casimir function and $k$ is some constant. Hence
the algebra-valued gradients $\nabla H_i$ satisfy the ``modified''
zero-curvature equations (\ref{modzce}). On the other hand, it
is not difficult to show, for the case of the Lie algebras
${\mathfrak{g}}$ described in the Proposition the
algebra-valued gradients of the Casimir functions are formal
power series. In particular, if the coadjoint representation
is equivalent to the adjoint one, they are proportional to a power of the
generic element of the dual space $L$, which in such a case is an
infinite linear combinations of the basic elements of
${\mathfrak{g}}$. On the other hand, due to the
assumption that all $H_i$ are finite polynomials, their
algebra-valued gradients are {\it finite} linear combinations of
the basic elements of the Lie algebra
${\mathfrak{g}}$. Hence the corresponding modified
zero-curvature equations are satisfied if and only if the
corresponding coefficient  $k$ in these equations is equal to
zero, i.e. when they pass to the usual zero-curvature conditions.
This proves the Theorem.

{\it Remark 5.}  Using the same arguments it is possible to also show
that a similar theorem holds true for more complicated
infinite-dimensional Lie algebras, e.g for the algebras of the
type $A_{\infty}$, $C_{\infty}$,  $D_{\infty}$ etc. In this case
instead of the condition that $H_i$ are finite polynomials  one
may require less rigid condition.

\subsection{Doubles, $R$-operators and negative flows of soliton hierarchies}
Now, using the results of previous section it is possible to
construct hierarchy of integrable equations in partial derivatives
admitting zero-curvature representations.

Indeed, we have obtained using classical double
$\mathcal{G}$, a large commutative algebra on the
quotients of $\mathcal{G}^*$. On the other hand, due
to the Theorem \ref{zceprop} in order to obtain
zero-curvature conditions as a consequence of the corresponding
commutative flows we will require that $\mathfrak{g}$
be infinite-dimensional and possesses infinitely many Casimir
invariants that will produce infinitely many commuting flows. In
such a case we will obtain, as a consequence of the general
Theorem \ref{zceprop}, the following Proposition:

\begin{tve}
Let  algebra $\mathfrak{g}$ be infinite-dimensional.
Denote $\mathcal{G}$ its double. Let
$\mathcal{G}^*$ be the corresponding dual space. Let
$J$ be ideal in
$\mathcal{G}_{\mathcal{R}_-}$. Denote $\pi:
\mathcal{G} \rightarrow \mathcal{G}/
J $ the natural projection onto the quotient
algebra. Let $\pi^*: (\mathcal{G}/ J )^* \rightarrow\mathcal{G}^*
$ be  the dual map. If
the functions $I_{k,\e}(\pi^*(\mathcal{L}))$ on
$\mathcal{G}^*$  are finite polynomials then the
$\mathfrak{g}$-valued functions
$$V_{k,+}=
{R}_{+}{\nabla} I_{k,1}(\pi^*(\mathcal{L})),\
 V_{l,-}= {R}_{-}{\nabla}
I_{l,2}(\pi^*(\mathcal{L}))$$
 satisfy zero-curvature equation with
values in $\mathfrak{g}$:
\begin{equation}\label{zce1}
\frac{ \partial V_{k,\pm}}{\partial t^{\pm}_{l}} - \frac{ \partial
V_{l,\pm}}{\partial t^{\pm}_k} + [V_{k,\pm}, V_{l,\pm}]= 0,
\end{equation}
\begin{equation}\label{zce2}
\frac{ \partial V_{k,\pm}}{\partial t^{\mp}_{l}} - \frac{ \partial
V_{l,\mp}}{\partial t^{\pm}_k} + [V_{k,\pm}, V_{l,\mp}]= 0.
\end{equation}
\end{tve}

{\it Remark 6.} Note, that equations (\ref{zce1})--(\ref{zce2})
define three types of integrable hierarchies: two ``small''
hierarchies associated with the Lie subalgebras
$\mathfrak{g}_{R_{\pm}}$ defined by equations
(\ref{zce1}) and one ``large'' hierarchy associated with the whole
Lie algebra $\mathfrak{g}$, that include both types of
equations (\ref{zce1}) and (\ref{zce2}). Equations (\ref{zce2})
contain $U$-$V$ pair with the $U$-operators taking their values in $\mathfrak{g}_{R_{+}}$ and $V$-operator
taking the values in $\mathfrak{g}_{R_{-}}$.
They have an interpretation of the ``negative flows'' of the
integrable hierarchy associated with
$\mathfrak{g}_{R_{\pm}}$.

\subsection{Case of graded Lie algebras.}
In this subsection we will demonstrate how the described above
general scheme of production of $U$-$V$ pairs satisfying
zero-curvature equations works for the concrete Lie algebras. We
will concentrate on the simplest possible examples associated with
the graded Lie algebras.
\subsubsection{Quotients of double and invariant functions.}\label{sec431}
Let us now consider the example of $\mathbb{Z}$-graded algebras
and the quotient algebras of the corresponding double. By the definition of graded Lie algebras  we have that
$$\mathfrak{g}=\sum\limits_{j \in \mathbb{Z}} \mathfrak{g}_j,
 \  \ \ [ \mathfrak{g}_i, \mathfrak{g}_j]\subset \mathfrak{g}_{i+j}.$$
From this one obtains a decomposition
$\mathfrak{g}=\mathfrak{g}_+ +
\mathfrak{g}_-$, where
$\mathfrak{g}_+=\sum\limits_{j \geq 0}
\mathfrak{g}_j$,
$\mathfrak{g}_-=\sum\limits_{j < 0}
\mathfrak{g}_j$ are Lie subalgebras. Denote $P_{\pm}$
the projection operators onto the Lie subalgebras
$\mathfrak{g}_{\pm}$. Hence $R=P_+-P_-$ is a classical
$R$-operator \cite{RST1} . In a standard way \cite{RST1} one
obtains that $J_{+k}=\sum\limits_{j>k}
\mathfrak{g}_j$ and
$J_{-l}=\sum\limits_{j>l}
\mathfrak{g}_{-l}$ are ideals in
$\mathfrak{g}_{\pm}$ and in
$\mathfrak{g}_{R}= \mathfrak{g}_{+}\ominus
\mathfrak{g}_{-}$. Hence one can consider the quotient
algebra $\mathfrak{g}_R/(J_{+k}\ominus
J_{-l})$ and the quotient algebra of the corresponding
``double'': $\mathcal{G}_{\mathcal
R}/(J_{+k}\ominus J_{-l})$. The elements
of this quotient algebra have the following form:
$\left(%
\begin{array}{c}
  X_1^+ + X_1^{-'} \\
 X_2^- + X_2^{+'} \\
\end{array}%
\right)$ where $X_1^+\in \mathfrak{g}_+$, $X_1^{-'}\in
\sum\limits_{j =1}^l \mathfrak{g}_{-j}$, $X_2^-\in
\mathfrak{g}_-$, $X_2^{+'}\in \sum\limits_{j = 0}^k
\mathfrak{g}_{j}$.

The corresponding elements of the dual space
have the following explicit form:
$\mathcal{L}=\left(%
\begin{array}{c}
  L_1^+ + L_1^{-'} \\
 L_2^- + L_2^{+'} \\
\end{array}%
\right)$,  where $L_1^+\in \mathfrak{g}^*_+$,
$L_1^{-'}\in \sum\limits_{j = 1}^l
\mathfrak{g}^*_{-j}$, $L_2^-\in
\mathfrak{g}^*_-$, $L_2^{+'}\in \sum\limits_{j =0}^k
\mathfrak{g}^*_{j}$. Let us notice once more that the
corresponding components of the Lax operator $L_1=L_1^+ +L_1^{-'}$ and $L_2=L_2^- +L_2^{+'}$ are
semi-infinite (i.e., infinite only in one direction).

Let us assume that on $\mathfrak{g}$ there is an
invariant bilinear form $(\ ,\ )$ such that
$(\mathfrak{g}_{i},\mathfrak{g}_{j} ) \sim
\delta_{i+j,0}$. In this case one can identify the spaces
$\mathfrak{g}^*$ and $\mathfrak{g}$ and
construct the  second order Casimir functions by the following
formula:
$$I^0_{2,1}= \frac{1}{2}\sum\limits_{i\in \mathbb{Z}} (L_1^{(i)}, L_1^{(-i)}), \  I^0_{2,2}= \frac{1}{2}\sum\limits_{i\in \mathbb{Z}} (L_2^{(i)}, L_2^{(-i)}),$$
where $L_{1, 2}^{(\pm i)} \subset  \mathfrak{g}_{\mp i}$.
Note that on the quotient algebra described above all these
expressions are finite polynomials if the space
$\mathfrak{g}_{i}$ is finite-dimensional.

Let us consider several Examples.

\subsubsection{General non-abelian Toda  systems and graded Lie algebras}
Let us consider the above construction in the case $k=1$, $l=1$.
In this case we have:
$$I^0_{2,1}= \frac{1}{2} (L_1^{(0)}, L_1^{(0)}) + (L_1^{(1)}, L_1^{(-1)}) , \
  I^0_{2,2}= \frac{1}{2} (L_2^{(0)}, L_2^{(0)}) + (L_2^{(1)}, L_2^{(-1)}),$$
and the Lax matrix is: $\mathcal{L}=\left(%
\begin{array}{c}
  L_1^{+} + L_1^{(-1)} \\
 L_2^{-} + L_2^{(0)}+L_2^{(1)} \\
\end{array}%
\right)$, where $L_1^{+}=\sum\limits_{i=0}^{\infty} L_1^{(i)}$, $
L_2^{-}=\sum\limits_{i=1}^{\infty} L_1^{(-i)}$. Let us also note,
that $L_1^{(-1)}$ is a central element due to the fact that
$L_1^{(-1)} \in
(\mathfrak{g}_-/[\mathfrak{g}_-,\mathfrak{g}_-])^*$.

 The  $M$-operators corresponding to the above integrals $I^0_{2,1}$, $I^0_{2,2}$   have the following form:
$${M}^0_{2,1}=\mathcal{R}_+ \nabla I^0_{2,1}= \left(%
\begin{array}{c}
  \bar{L}_1^{(0)}+ \bar{L}_1^{(1)} \\
  \bar{L}_1^{(0)}+ \bar{L}_1^{(1)}  \\
\end{array}%
\right), \quad {M}^0_{2,2}=\mathcal{R}_+ \nabla I^0_{2,2}=-\left(%
\begin{array}{c}
   \bar{L}_2^{(-1)} \\
 \bar{L}_2^{(-1)}  \\
\end{array}%
\right), \text{ where }$$ $\bar{L}_1^{(0)}=\frac{1}{2}\nabla
({L}_1^{(0)},{L}_1^{(0)}) \in
\mathfrak{g}_{0}$, $\bar{L}_1^{(1)}=P_+\nabla
({L}_1^{(1)},{L}_1^{(-1)}) \in
\mathfrak{g}_{1}$, $\bar{L}_2^{(-1)}=P_-\nabla
({L}_2^{(1)},{L}_2^{(-1)}) \in
\mathfrak{g}_{-1}$.

Their components, namely,  the operators
\begin{equation}\label{uvtoda}
U=\bar{L}_1^{(0)}+ \bar{L}_1^{(1)},\quad V=\bar{L}_2^{(-1)}
\end{equation}
are
$U$-$V$ pairs of abelian and non-abelian Toda field equations, as we will show in a moment. Let
us also note, that $\bar{L}_1^{(1)}$ is a constant element
because $L_1^{(-1)}$ is a central element that can be identified with a constant.

Let us consider the  corresponding zero-curvature equation in the graded Lie algebra
$$\frac{\partial U}{ \partial t} - \frac{\partial V}{ \partial x}
+ [U,V]=0.$$ It yields the following equations for the homogeneous components
\begin{gather}\label{natzc}
 \frac{\partial \bar{L}_1^{(1)}}{\partial t}=0,\
\frac{\partial \bar{L}_1^{(0)}}{ \partial
t}=-[\bar{L}_1^{(1)}, \bar{L}_2^{(-1)}], \ \frac{\partial
\bar{L}_2^{(-1)}}{
\partial x}=[\bar{L}_1^{(0)}, \bar{L}_2^{(-1)}].
\end{gather}
The first of these equations is satisfied automatically, using the fact
that $\bar{L}_1^{(1)}$ is obtained from the central element $L_1^{(-1)}$
and we can put $\bar{L}_1^{(1)}=C^{(1)}=const$.  Let us
solve the last equation of \eqref{natzc}. Due to the grading  it is easy to see that
$\mathfrak{g}_0\subset \mathfrak{g}$ is a subalgebra. Denote $G_0$ the
corresponding Lie group.
 Let $g_0 \in G_0$. By
 direct verification one can show that the substitution
$$\bar{L}_2^{(-1)}= g_0 C^{(-1)} g_0^{-1}, \
\bar{L}_1^{(0)}= (\partial_x g_0)g_0^{-1},\quad
g_0=g_0(x,t)$$ where $C^{(-1)}$ is a constant element of the space
$\mathfrak{g}_{-1}$, solves the last of the three
equations (\ref{natzc}). The second of the equations (\ref{natzc})
takes after the substitution of this solution the following form:
\begin{gather}\label{nat}
\partial_t \left((\partial_x g_0)g_0^{-1}\right)
=-[C^{(1)}, g_0 C^{(-1)} g_0^{-1}].
\end{gather}
Equation (\ref{nat}) is the so-called non-abelian Toda field
equations \cite{Raz}.
\subsubsection{Loop algebras and the standard Toda system}
 The main example  of the above construction is connected with loop algebras. Let $\mathfrak{a}$ be the Lie algebra of a simple Lie group $G$.
  Denote $\mathfrak{g}=\mathfrak{a}\otimes Pol(\l,\l^{-1})$ the loop
 algebra. Assume that  $\mathfrak{a}$ is equipped with an automorphism $\sigma: \mathfrak{a}\to \mathfrak{a}$ of order $p$. One has a natural decomposition \cite{Kac}
 $$\mathfrak{a}=\sum\limits_{i=0}^{p-1}\mathfrak{a}_i$$ such that $$\mathfrak{a}_k=\{X\in \mathfrak{a}| \sigma(X)=e^{\frac{2\pi i k}{p}} X\}.$$
In particular $\mathfrak{a}_0$ is the subalgebra
stable under the action of  automorphism $\sigma$.

Let us extend this grading to the loop space
$\mathfrak{g}$ prescribing by the definition
\begin{eqnarray}
&&
\mathrm{{deg}\,} \l=p
\nonumber\\
&&
\mathrm{{deg}\,} X\otimes q(\l)= \mathrm{deg }\,X +
\mathrm{deg }\,q(\l).
\nonumber
\end{eqnarray}
In this case we obtain
$$\mathfrak{g}_{j}= \{X(\l) \in \mathfrak{a}\otimes
Pol(\l,\l^{-1})\,|\, \mathrm{deg}\, X(\l) =j\}.$$ In particular  $\mathfrak{g}_{0}=\mathfrak{a}_0$. Hence the corresponding group $G_0\subset G$ coincides with the Lie group of the Lie subalgebra $\mathfrak{a}_0$.
The equation (\ref{nat}) is written for the generic element of this group.

Let us consider the most interesting example of such a situation
corresponding to the case of abelian $G_0$. Let
$\mathfrak{g}$ be a loop
 algebra with the principal grading. In more details, $\sigma:\mathfrak{a}\to\mathfrak{a}$ is a Coxeter automorphism. Denote $h$ the Coxeter number of $\mathfrak{a}$. We have the corresponding $\mathbb Z_h$-grading of $\mathfrak{a}$
 $$\mathfrak{a}=\sum\limits_{i=0}^{h-1}\mathfrak{a}_i.$$
The subalgebra  $\mathfrak{a}_0$ coincides with the Cartan subalgebra
$$
\mathfrak{h}=\mathrm{Span}_{\mathbb{C}}\{ H_{\a_i}| i\in
 1,\dots , \mathrm{rank}\, \mathfrak{g} \}.$$
 Moreover
  $$\mathfrak{a}_k=\mathrm{Span}_{\mathbb{C}}\{ X_{\a}| \a \in \Delta, |\a|=k\ \mathrm{mod}\ h
  \}.$$ Here  $ H_{\a_i}$, $X_{\a}$ is a Cartan-Weil basis of
  $\mathfrak{a}$, $\Delta$ is the set of all roots,  and
$|\a|$ stands for the height of root. In particular $H_{\a_i}=[X_{\a_i},X_{-\a_i}]$,
  where $\a_i$ are simple roots.

Let us describe more explicitly the subspaces
$\mathfrak{g}_i$.
 By definition we have:
$$\mathfrak{g}_0=\mathfrak{h}, \  \mathfrak{g}_k =\sum_{|\a|= k} \mathfrak{a}_{\a} +
\l \sum_{|\a|= h-k} \mathfrak{a}_{-\a}, \quad k \in \overline{1,h-1},$$
where $ \mathfrak{a}_{\a}= Span_{\mathbb{C}}\{ X_{\a}\}$. The other graded subspaces are:
$$\mathfrak{g}_{k+nh}=\l^{n}\mathfrak{g}_k, \text{ where } k\in \overline{1,h-1}.$$
Let us consider the corresponding $U$-$V$-pair
(\ref{uvtoda})
$$
U=(\partial_x g_0)g_0^{-1}+ C^{(1)},\quad V=g_0 C^{(-1)} g_0^{-1}.
$$
In this case the group $G_0$ is abelian and coincides with the
Cartan subgroup. Because of this it is easy to parametrize the
element $g_0$ in the following way:
$g_0=\exp{\sum\limits_{i=1}^{\mathrm{rank} \mathfrak{a}}\phi_i
H_{\a_i}}$ and obtain that
\begin{gather*}U=\sum\limits_{i=1}^{{\rm rank} \mathfrak{a}}\partial_x \phi_i H_{\a_i} + \sum\limits_{\a_i \in P} c^{(1)}_{\a_i} X_{\a_i}+ \l c^{(1)}_{-\theta} X_{- \theta}\\
 V=\sum\limits_{\a_i \in P} c^{(-1)}_{\a_i} e^{-\a_i(\phi)} X_{-\a_i}+ \l^{-1} c^{(-1)}_{\theta} e^{-\theta(\phi)}X_{\theta},
 \end{gather*}
where  $P$ is the set of simple roots, $\theta$ is the highest root
and $H_{\a_i}$ is the basic element in the Cartan subalgebra
corresponding to the simple root $\a_i$.

It is easy to recognize in this $U$-$V$ pair  the $U$-$V$ pair of
finite-component Toda field equation \cite{Mikh2}. The
corresponding equations (\ref{nat}) have the following form:
\begin{gather}\label{at}
\partial_t \partial_x \phi_i
=c^{(1)}_{\a_i}c^{(-1)}_{-\a_i} e^{-\a_i(\phi)}+a_i
c^{(1)}_{-\theta}c^{(-1)}_{\theta}e^{\theta(\phi)},
\end{gather}
where $\phi= \sum\limits_{i=1}^{{\rm rank} \mathfrak{a}}\phi_i H_{\a_i}$
and the
 constants $a_i$ are defined from the decomposition of $H_{\theta}=[X_{\theta}, X_{-\theta}]$
 $$H_{\theta}=\sum\limits_{i=1}^{\mathrm{rank} \mathfrak{a}} a_i
H_{\a_i}$$

It is easy to see from the very form of the equations (\ref{at})
that the coefficients $c^{(-1)}_{-\a_i}$, $c^{(1)}_{\a_i}$,
$c^{(-1)}_{\theta}$, $c^{(1)}_{-\theta}$   are
redundant and, if non-zero, they can be  eliminated from the equations by a rescaling.\\

\subsubsection{ Infinite-component Toda system}
Let now
$\mathfrak{g}=gl((\infty))$. Recall that this is the Lie algebra of infinite matrices
$$
M=\left(M_{ij}\right)_{i, \, j\in\mathbb Z}, \quad M_{ij}=0 \quad \mbox{for}\quad |i-j|>>1.
$$ This situation can be considered as the
$n\rightarrow \infty$ limit of the case $\mathfrak{g}=gl(n)$ of
the previous section. However it deserves more careful considerations.

The basis in the algebra $gl((\infty))$ consists of the elements
$X_{ij}$, $i,j \in \mathbb{Z}$ with the standard commutation
relations:
$$[X_{ij},X_{kl}]=\d_{kj}X_{il}-\d_{il}X_{kj}.$$
In terms of this basis we have the following graded subspaces
of the natural $\mathbb{Z}$-grading:
$$\mathfrak{g}_k=Span_{\mathbb{C}}\{ X_{ij}\,|\,  j-i=k
  \}.$$
On $gl((\infty))$ there exists a natural invariant bilinear form $(\ ,\ )$
such that
$$(X_{ij},X_{kl})=\d_{kj}\d_{il}.$$ Using this form one identifies
$\mathfrak{g}^*$ with $\mathfrak{g}$ so
that $\mathfrak{g}^*_k=\mathfrak{g}_{-k}$.

Let us consider  the classical double of $gl((\infty))$. We apply
the construction of Section \ref{sec431} to the corresponding dual
space and its quotient spaces with respect to the ideals of the
form $J_{+k}$ and $J_{-l}$, $k$, $l$ are
given positive integers.  The elements of the dual spaces to these
quotients
$$
\mathcal{L}=\left(%
\begin{array}{c}
  L_1^+ + L_1^{-'} \\
 L_2^- + L_2^{+'} \\
\end{array}%
\right)\in \left[\mathcal{G}_{\mathcal
R}/(J_{+k}\ominus J_{-l})\right]^*
$$
 have the following explicit form
 $$
 L_1^{+}=\sum\limits_{i=0}^{\infty} L_1^{(i)}, \quad
L_2^{-}=\sum\limits_{i=1}^{\infty} L_2^{(-i)},\quad L_1^{-'}=
\sum\limits_{j = 1}^l L_1^{(-i)},\quad L_2^{+'}=\sum\limits_{j =0}^k
L_1^{(i)},
$$
$L_s^{(i)} \in gl((\infty))_{-i}$ ($s = 1,2$).
In particular the Lax matrix ${\cal L}$ of the infinite-component Toda system
will correspond to the case $k=l=1$. So we will consider only this case in sequel.

We will denote the natural basis in the dual space to $gl((\infty))$ by the same symbols $X_{ij}$.
In this basis the Lax operator ${\cal L}$ can be described by the coordinates $l_1^{(m)}(i)$, $l_2^{(m)}(i)$
in the following manner
$$
L_s^{(m)}=\sum\limits_{i \in
\mathbb{Z}} l^{(m)}_{s}(i-m)X_{i,i-m}, \quad s=1, \, 2.
$$
As above the coordinates $l_1^{(-1)}(i)$ are Casimirs of the
Lie--Poisson bracket on the dual to the quotient
$\mathcal{G}_{\mathcal R}/(J_{+1}\ominus
J_{-1})$. So we can  put them to be equal to
constants, $l_1^{(-1)}(i)=c_i$. Thus
$$
L_1^{(-1)}=\sum_{i\in\mathbb Z} c_i X_{i-1, i}.
$$

Using the invariant form $(~,~)$ on $gl((\infty))$ we obtain two quadratic
Hamiltonians $I^0_{2,1}$, $I^0_{2,2}$ on the double of $gl((\infty))$  having the following
explicit form on the quotient space under consideration
$$I^0_{2,s}= \frac{1}{2}\sum\limits_{i \in
\mathbb{Z}} \left(l_{s}^{(0)}(i)\right)^2 + \sum\limits_{i \in \mathbb{Z}}
l_{s}^{(1)}(i)l_{s}^{(-1)}(i), \ s\in 1,2.$$
The flows generated by these Hamiltonians are written in the form
$$
\frac{\partial {\cal L}}{\partial t_s} =\left[ \tilde M_{2,s}^0, {\cal L}\right], \quad s=1, \, 2
$$
where
the $M$-operators with
values in the double of $gl((\infty))$  are
$$\widetilde{M}^0_{2,1}=\mathcal{R}_+\widetilde{\nabla} I^0_{2,1}=\left(%
\begin{array}{c}
 {L}_1^{(0)}+ {L}_1^{(-1)} \\
  {L}_1^{(0)}+ {L}_1^{(-1)} \\
\end{array}%
\right),\  \widetilde{M}^0_{2,2}=\mathcal{R}_+\widetilde{\nabla} I^0_{2,2}=\left(%
\begin{array}{c}
{L}_2^{(1)} \\
{L}_2^{(1)} \\
\end{array}%
\right).
$$
Despite of the fact that in this case $\mathfrak{g}$
is not a loop algebra  and functions $I^0_{2,s}$ are not finite
polynomials, one can prove, in a similar way to the proof of the
Theorem \ref{zceprop}, that the corresponding $M$-operators
satisfy zero-curvature condition. Hence, one can write
 the  following  $gl((\infty))$-valued  $U$-$V$-pair  satisfying zero-curvature equation:
$$U= {L}_1^{(0)}+ {L}_1^{(-1)},\quad V= {L}_2^{(1)}, \text{ where } L_s^{(i)} \in gl((\infty))_{-i}, \quad s=1, \, 2.$$
It yields the following equations:
\begin{gather}\label{toda2}
\partial_x v_i=v_i(u_{i+1}-u_i),\
\partial_t u_i=c_{i-1} v_{i-1}-c_iv_{i}, \  i \in \mathbb{Z},
\end{gather}
where
$$
u_i\equiv l_{1}^{(0)}(i), \quad v_i\equiv l_{2}^{(1)}(i),\quad  x=t_1, \quad t=t_2.
$$
By
the substitution  $u_i=\partial_x \phi_i$,
$v_i=e^{\phi_{i+1}-\phi_i}$ the equations (\ref{toda2}) reduce to the
usual infinite-component Toda equations \cite{Ueno}:
\begin{gather}\label{toda3}
\partial^2_{xt}\phi_i=c_{i-1}
e^{\phi_{i}-\phi_{i-1}}-c_ie^{\phi_{i+1}-\phi_i}, \  i \in
\mathbb{Z}.
\end{gather}
As above the constants $c_i$, if non-zero, can be eliminated by a rescaling.

 {\it Remark 7.}  Note that the equation (\ref{toda3}) does not coincide in its form with the
$gl(n)$-equations (\ref{at}) because in this subsection we have
worked with another basis in
 Cartan subalgebra: $H_i\equiv X_{ii}$ instead of
 $H_{\a_i}=X_{ii}-X_{i-1i-1}$.

\subsubsection{ Lie--Poisson bracket for the infinite-component Toda system}
In this subsection for the purpose of illustration we will
explicitly describe the $R$-matrix Lie--Poisson bracket for the case of
the Lie algebra $\mathfrak{g}=gl((\infty))$, for its
double and  for the  $R$-operator corresponding to the natural
Adler-Kostant-Symes decomposition used in the previous
subsection. We will start from the  Lie brackets first and then
use the fact that the Lie--Poisson brackets of the coordinate functions can be easily recovered from the Lie brackets of the basic elements.

Let us, for the purpose of convenience introduce the following
basis in the algebra $gl((\infty))$:
$$X^{(i)}(m)\equiv X_{m, i+m}, \quad i, \, m\in\mathbb Z.$$ The commutation relations in this
basis acquires the following form:
$$[X^{(i)}(m),X^{(j)}(n)]=\d_{i+n-m,0} X^{(i+j)}(m) - \d_{m-n+j,0}
X^{(i+j)}(n).$$ The $R$-operator in the case under consideration has the
following form: $R=P_+-P_-$, where $P_{\pm}$ are the projection
operators onto the Lie subalgebras generated by $X^{(i)}(m)$,
$i\geq 0$ and $X^{(j)}(n)$, $j<0$ respectively.

The $R$-bracket on $gl((\infty))$ can be written as
follows:
$$[X^{(i)}(m),X^{(j)}(n)]_R=2(1-\sigma(i)-\sigma(j))\bigl(\d_{i+n-m,0} X^{(i+j)}(m) - \d_{m-n+j,0}
X^{(i+j)}(n)\bigr),$$ where $\sigma(i)=1$ if $i<0$, $\sigma(i)=0$ if $i\geq
0$.

For the double of $gl((\infty))$, namely for the direct sum
$gl((\infty))\oplus gl((\infty))$ we obtain the following
$\mathcal{R}$-bracket written for the basic elements
$X_s^{(i)}(m)$, $s\in 1,2$:
\begin{subequations}\label{gindcr}
\begin{equation}
[X_1^{(i)}(m),X_1^{(j)}(n)]_R=2(1-\sigma(i)-\sigma(j))\bigl(\d_{i+n-m,0}
X_1^{(i+j)}(m) - \d_{m-n+j,0} X_1^{(i+j)}(n)\bigr),
\end{equation}
\begin{equation}
[X_2^{(i)}(m),X_2^{(j)}(n)]_R=2(\sigma(i)+\sigma(j)-1)\bigl(\d_{i+n-m,0}
X_2^{(i+j)}(m) - \d_{m-n+j,0} X_2^{(i+j)}(n)\bigr),
\end{equation}
\begin{multline}
[X_1^{(i)}(m),X_2^{(j)}(n)]_R=2(\sigma(i)-1)\bigl(\d_{i+n-m,0}
X_1^{(i+j)}(m) - \d_{m-n+j,0}
X_1^{(i+j)}(n)\bigr)+\\+2\sigma(j)\bigl(\d_{i+n-m,0} X_2^{(i+j)}(m) -
\d_{m-n+j,0} X_2^{(i+j)}(n)\bigr).
\end{multline}
\end{subequations}
The Lie--Poisson brackets for the coordinate functions
$l^{(i)}_s(m)$ readily follow from the commutation relations
(\ref{gindcr})
\begin{subequations}
\begin{equation}
\{l_1^{(i)}(m),l_1^{(j)}(n)\}_R=2(1-\sigma(i)-\sigma(j))\bigl(\d_{i+n-m,0}
l_1^{(i+j)}(m) - \d_{m-n+j,0} l_1^{(i+j)}(n)\bigr),
\end{equation}
\begin{equation}
\{l_2^{(i)}(m),l_2^{(j)}(n)\}_R=2(\sigma(i)+\sigma(j)-1)\bigl(\d_{i+n-m,0}
l_2^{(i+j)}(m) - \d_{m-n+j,0} l_2^{(i+j)}(n)\bigr),
\end{equation}
\begin{multline}
\{l_1^{(i)}(m),l_2^{(j)}(n)\}_R=2(\sigma(i)-1)\bigl(\d_{i+n-m,0}
l_1^{(i+j)}(m) - \d_{m-n+j,0}
l_1^{(i+j)}(n)\bigr)+\\+2\sigma(j)\bigl(\d_{i+n-m,0} l_2^{(i+j)}(m) -
\d_{m-n+j,0} l_2^{(i+j)}(n)\bigr).
\end{multline}
\end{subequations}
The Lie--Poisson brackets of the Toda system are obtained by
 putting in these relations $l_{1}^{(-1)}(i)=c_i$,
$l_{1}^{(k)}(i)=0$, $k<-1$, $l_{2}^{(j)}(i)=0$, $j>1$. These
brackets coincide with the first Poisson structure of the 2D Toda
hierarchy found in \cite{GCar}.
\section{ Quadratic and cubic Poisson structures on  Double}
In this section we will discuss the prolongation of the second and
third degree Poisson brackets from $\mathfrak{g}$ to
its classical double $\mathcal{G}$ and the consistency
of the corresponding brackets.
\subsection{ Quadratic  Poisson structure}
It is known that for some classical $R$-operators
on $\mathfrak{g}$ it is possible to define, besides the linear
$R$-bracket, also a second degree Poisson bracket important
in the theory of classical integrable systems.

Hereafter we assume existence of an identification between $\mathfrak{g}$ and $\mathfrak{g}^*$. Moreover we assume that the Lie algebra $\mathfrak{g}$  possesses also the structure
of an associative algebra.
The following Theorem holds true \cite{RaOe}, \cite{LiPa}:
\begin{teo}\label{rolp}
Let the classical $R$-operator and its skew-symmetric part
$\frac{1}{2}(R-R^*)$
 satisfy modified classical Yang-Baxter
equation on $\mathfrak{g}$. Then

(i) The formula
\begin{equation}\label{quadr}
\{F_1,F_2\}_2=\langle L, [R(L\nabla F_1+ \nabla F_1 L),\nabla
F_2]\rangle- \langle L, [R(L\nabla F_2+ \nabla F_2 L),\nabla
F_1]\rangle
\end{equation}
defines a Poisson bracket on $\mathfrak{g}$.

(ii) The Casimir functions of $\mathfrak{g}$ mutually
commute with respect to the  brackets (\ref{quadr}).

(iii) The Hamiltonian equations with respect to the Casimir
functions $I_k$ of $\mathfrak{g}$ are written in the
Lax form:
$$\frac{d L}{d t_k}=[R(L\nabla I_k+ \nabla I_k L),L]$$

(iv) The Poisson brackets (\ref{quadr}) and (\ref{rpb}) are
compatible.
\end{teo}

Recall that two Poisson brackets $\{~, ~\}_1$ and $\{~, ~\}_2$ on the same space are called \emph{compatible} if an arbitrary linear combination
$$
a_1 \{~, ~\}_1+a_2\{~, ~\}_2
$$
is again a Poisson bracket.

It occurred that the Theorem \ref{rolp} can be extended to the
double of $\mathfrak{g}$:

\begin{teo}\label{dus}
Let the classical $R$-operator and its skew-symmetric part
$\frac{1}{2}(R-R^*)$
 satisfy modified classical Yang-Baxter
equation on $\mathfrak{g}$. Then

(i) The formula
\begin{equation}\label{quadrd}
\{F_1(\mathcal{L}),F_2(\mathcal{L})\}_2=\langle \mathcal{L},
[\mathcal{R}(\mathcal{L}\widetilde{\nabla} F_1+ \widetilde{\nabla}
F_1 \mathcal{L}),\nabla F_2]\rangle- \langle \mathcal{L},
[\mathcal{R}(\mathcal{L}\widetilde{\nabla} F_2+ \widetilde{\nabla}
F_2 \mathcal{L}),\nabla F_1]\rangle
\end{equation}
defines a Poisson bracket on $\mathcal{G}$.

(ii) The Casimir functions of $\mathcal{G}$ mutually
commute with respect to the  brackets (\ref{quadrd}).

(iii) The Hamiltonian equations with respect to the Casimir
functions $I_{k,\e}$ of   $\mathcal{G}$  are written
in the Lax form:
$$\frac{d \mathcal{L}}{d t^{\e}_k}=[R(\mathcal{L}\widetilde{\nabla} I_{k,\e}+ \widetilde{\nabla} I_{k,\e} \mathcal{L}),\mathcal{L}], \quad \epsilon=1, \, 2.$$

(iv) The Poisson brackets (\ref{quadrd}) and (\ref{pbd}) are
compatible.
\end{teo}

{\it Remark 8.} The Lie--Poisson bracket (\ref{quadrd}) can be
written in terms of the operators $R$, $R_{\pm}$ more
explicitly as follows:
\begin{gather*}
\{F_1(L_1,L_2),F_2(L_1,L_2)\}_2=\langle L_1,
[{R}({L}_1\nabla_1F_1+\nabla_1 F_1 {L}_1)- R_-({L}_2\nabla_2 F_1
+\nabla_2 F_1 {L}_2),\nabla_1 F_2]\rangle+\\+ \langle L_2,
[{R}_+({L}_1\nabla_1 F_1+ \nabla_1 F_1 {L}_1)- {R}({L}_2\nabla_2
F_1+ \nabla_2 F_1 {L}_2),\nabla_2 F_2]\rangle -\\- \langle L_1,
[{R}({L}_1\nabla_1F_2+\nabla_1 F_2 {L}_1)- R_-({L}_2\nabla_2 F_2
+\nabla_2 F_2 {L}_2),\nabla_1 F_1]\rangle-\\- \langle L_2,
[{R}_+({L}_1\nabla_1 F_2+ \nabla_1 F_2 {L}_1)- {R}({L}_2\nabla_2
F_2+ \nabla_2 F_2 {L}_2),\nabla_2 F_1]\rangle.
\end{gather*}
{\it Proof.}
 Let us at first note that if $\mathfrak{g}$ is an associative algebra then
 $\mathcal{G}$ is also an associative algebra with respect to the
 natural structure of direct sum of associative algebras.
 In order to prove the theorem it will suffice to apply the Theorem \ref{rolp}. Namely, to derive
from the assumptions about the classical $R$-operator on $\mathfrak{g}$ that the classical
$R$-operator $\mathcal{R}$ and its skew-symmetric part
$\frac{1}{2}(\mathcal{R}-\mathcal{R}^*)$
 satisfy modified classical Yang-Baxter
equation on the double $\mathcal{G}$. The first part of the
statement follows automatically from the results of \cite{RST2}.
It remains to show that $\frac{1}{2}(\mathcal{R}-\mathcal{R}^*)$
satisfies modified classical Yang-Baxter equation on
$\mathcal{G}$. We will do this by direct calculation.
We have that $$\mathcal{A}\equiv
\frac{1}{2}(\mathcal{R}-\mathcal{R}^*)=\frac12
\begin{pmatrix}
  A & -S \\
  S & -A \\
\end{pmatrix},\quad
A=R-R^*, \quad S= R+R^*.$$
Substituting
this expression  into the modified classical Yang-Baxter equation
one obtains that it is equivalent to the following three
conditions
\begin{subequations}\label{rdid}
\begin{equation}\label{rdid1}
A([A(X),Y]+ [X,A(Y)])-[A(X),A(Y)]=4[X,Y],\ \ \  \forall X,Y \in
\mathfrak{g},
\end{equation}
\begin{equation}
-S([S(X),Y])- A([X,S(Y)])+[A(X),S(Y)]=0,\ \ \  \forall X,Y \in
\mathfrak{g},
\end{equation}
\begin{equation}
S([A(X),Y]+ [X,A(Y)])-[S(X),S(Y)]=0,\ \ \  \forall X,Y \in
\mathfrak{g}.
\end{equation}
\end{subequations}
The equation (\ref{rdid1}) follows from the very
condition of the theorem. Besides, using the definition of
the operators $A$ and $S$ it is straightforward to show that the
conditions (\ref{rdid}) are equivalent to the following three
equations
\begin{subequations}\label{rdidr}
\begin{equation}\label{rdidr1}
R([R(X),Y]+ [X,R(Y)])-[R(X),R(Y)]=[X,Y],\ \ \  \forall X,Y \in
\mathfrak{g},
\end{equation}
\begin{equation}\label{rdidr2}
R^*([R^*(X),Y])- R^*([X,R(Y)])+[R^*(X),R(Y)]=[X,Y],\ \ \  \forall
X,Y \in \mathfrak{g},
\end{equation}
\begin{equation}\label{rdidr3}
R([R^*(X),Y]+ [X,R^*(Y)])-[R^*(X),R^*(Y)]=-[X,Y],\ \ \  \forall
X,Y \in \mathfrak{g}.
\end{equation}
\end{subequations}
The equation (\ref{rdidr1}) is  a modified classical Yang-Baxter
equation for $R$. The equation (\ref{rdidr2}) is derived using
classical modified Yang-Baxter equation (\ref{rdidr1}) and  the
existence of the non-degenerate invariant form on
$\mathfrak{g}$. Finally the identity (\ref{rdidr3}) is
derived using the modified Yang-Baxter equation for $A$.

Theorem is proven.

{\it Remark 9.} Note that the Theorem \ref{dus} means that the
quadratic Poisson structure, whenever exists on
$\mathfrak{g}$ can be always extended to its double
$\mathcal{G}$. In particular such an extension exists for skew-symmetric
$R$-operators on $\mathfrak{g}$ as the corresponding
operator $\mathcal{R}$ on $\mathcal{G}$ is also
skew-symmetric.

\subsection{ Cubic  Poisson structure}
In this subsection we will describe the cubic Poisson structure on
$\mathfrak{g}$ and its prolongation to
$\mathcal{G}$. We will use  the following Theorem
\cite{RaOe}, \cite{LiPa} valid under the same assumptions about the Lie algebra $\mathfrak{g}$.
\begin{teo}\label{dcb}
Let  $R$ be the classical $R$-operator. Then

(i) The following formula:
\begin{equation}\label{cub}
\{F_1(L),F_2(L)\}_3=\langle L, [R(L\nabla F_1L+ L\nabla F_1
L),\nabla F_2]\rangle- \langle L, [R(L\nabla F_2L+ L\nabla F_2
L),\nabla F_1]\rangle
\end{equation}
defines a Poisson bracket on $\mathfrak{g}$.

(ii) The Casimir functions of $\mathfrak{g}$ mutually
commute with respect to the  brackets (\ref{cub}).

(iii) The Hamiltonian equations with respect to the Casimir
functions $I_k$ of $\mathfrak{g}$ are written in the
Lax form:
$$\frac{d L}{d t_k}=[R(L\nabla I_k L),L]$$

(iv) If the skew-symmetric part $\frac{1}{2}(R-R^*)$ of the
operator $R$
 satisfies modified classical Yang-Baxter
equation on $\mathfrak{g}$. Then the Poisson brackets
 (\ref{cub}), (\ref{quadr}), and (\ref{rpb}) are compatible.
\end{teo}

A similar statement holds true also for the
classical double of $\mathfrak{g}$.

\begin{nas}\label{cus}
Let  $R$ be the classical $R$-operator. Then

(i) the following formula:
\begin{equation}\label{cubd}
\{F_1(\mathcal{L}),F_2(\mathcal{L})  \}_3=\langle \mathcal{L},
[\mathcal{R}(\mathcal{L}\widetilde{\nabla} F_1\mathcal{L}+
\mathcal{L}\widetilde{\nabla} F_1 \mathcal{L}),\widetilde{\nabla}
F_2]\rangle- \langle \mathcal{L},
[\mathcal{R}(\mathcal{L}\widetilde{\nabla} F_2\mathcal{L}+
\mathcal{L}\widetilde{\nabla} F_2 \mathcal{L}),\widetilde{\nabla}
F_1]\rangle
\end{equation}
defines Poisson bracket on $\mathcal{G}$.

(ii) The Casimir functions of $\mathcal{G}$ mutually
commute with respect to the  brackets (\ref{quadrd}).

(iii) The Hamiltonian equations with respect to the Casimir
functions $I_{k,\e}$ of   $\mathcal{G}$  are written
in the Lax form:
$$\frac{d \mathcal{L}}{d t^{\e}_k}=[R(\mathcal{L}\widetilde{\nabla} I_{k,\e}\mathcal{L}),\mathcal{L}].$$

(iv) if the skew-symmetric part $\frac{1}{2}(R-R^*)$ of the
operator $R$
 satisfies modified classical Yang-Baxter
equation on $\mathfrak{g}$  then the Poisson brackets
 (\ref{cubd}), (\ref{quadrd}), and (\ref{rpbd}) are compatible.
\end{nas}

{\it Remark 10.} The Lie--Poisson bracket (\ref{cubd}) can be
written in terms of the operators $R$, $R_{\pm}$ more
explicitly as follows:
\begin{multline*}
\{F_1(L_1,L_2),F_2(L_1,L_2)\}_3=\\=\langle L_1,
[{R}({L}_1\nabla_1F_1{L}_1)- R_-({L}_2\nabla_2 F_1{L}_2),\nabla_1
F_2]\rangle+ \langle L_2, [{R}_+({L}_1\nabla_1 F_1 {L}_1)-
{R}({L}_2\nabla_2 F_1{L}_2),\nabla_2 F_2]\rangle -\\- \langle L_1,
[{R}({L}_1\nabla_1F_2{L}_1 )- R_-({L}_2\nabla_2 F_2{L}_2),\nabla_1
F_1]\rangle- \langle L_2, [{R}_+({L}_1\nabla_1 F_2{L}_1)-
{R}({L}_2\nabla_2 F_2{L}_2),\nabla_2 F_1]\rangle.
\end{multline*}

{\it Proof.} In order to prove the Corollary, let us first
observe that, using the identification between $\mathfrak{g}$ and
$\mathfrak{g}^*$ as linear spaces and
$\mathfrak{g}$-modules one can also identify the
spaces $\mathcal{G}$ and  $\mathcal{G}^*$
as linear spaces and $\mathcal{G}$-modules. As it was explained above the double $\mathcal{G}$ of an associative algebra also
possesses a structure of an associative algebra. Now in order to prove
the Corollary it suffices to apply the Theorem \ref{dcb} to the
classical double $\mathcal{G}$ and take into account
that, as it was proven in the Theorem \ref{dus}, the antisymmetric part
$\frac{1}{2}(\mathcal{R}-\mathcal{R}^*)$
 satisfies the modified classical Yang-Baxter
equation on $\mathcal{G}$ if $\frac{1}{2}({R}-{R}^*)$
satisfies modified classical Yang-Baxter equation on
$\mathfrak{g}$. This proves the
Corollary.\\

{\it Remark 11.} Note, that the second and third degree Poisson
structures exist and are compatible with the linear Poisson
structure for all associative algebras (provided that $R$ and
$\frac{1}{2}(R-R^*)$ satisfy the modified Yang-Baxter equation).
They produce commuting hamiltonian flows written in the Lax form.
Nevertheless their usage in the soliton theory is not as
straightforward and universal as the usage of the linear Poisson
$R$-bracket. Indeed, in order to obtain the phase space of the
soliton equation under consideration one has, as it was explained
above, to restrict oneself to certain linear subspaces coinciding
with the quotient algebras of the corresponding linear
$R$-bracket, i.e., to Poisson subspaces with respect to the linear
bracket. On the other hand, these linear subspaces in general are
not Poisson subspaces of the quadratic and cubic brackets.
Moreover, in order to restrict these brackets to the corresponding
 subspaces, one has to apply additionally Dirac's
reduction. For the case of the infinite-component Toda system this
procedure was considered   in \cite{GCar}.

At the end of this section let us finally explain why second and
third order Poisson structures exist and are compatible with the
linear Poisson structure  for the case of infinite-component Toda
system (i.e. before the restriction onto the Poisson subspaces of
linear bracket). As far as infinite-component Toda model is
connected with the double of $gl((\infty))$ we need to prove the
following:
\begin{tve}
On the double of $gl((\infty))$ equipped with the
Adler-Kostant-Symmes $R$-operator $R=P^+ - P^- $, where $P^+$ and
$P^-$ are projection operators onto the algebra of upper
triangular and strictly lower triangular matrices, exist quadratic
and cubic Poisson structures compatible with linear Poisson
brackets and given by the formulas (\ref{quadrd}) and (\ref{cubd})
correspondingly.
\end{tve}

{\it Proof.} To prove the proposition we note that the double of
$gl((\infty))$ is evidently an associative algebra. Hence in order
to  apply the Theorem \ref{dus}  and Corollary \ref{cus} for the
case at hand we have to show that $\frac{1}{2}(R-R^*)$ satisfies
classical Yang-Baxter equation on $gl((\infty))$. In our case we
have that: $R=P^+ - P^- $, where $P^+$ and $P^-$ are projection
operators onto the algebra of upper triangular and strictly lower
triangular matrices respectively. Let us also note that this
$R$-operator can be written as $R=P_+ +P_0-P_-$ where $P_+$
is the projection operator onto the Lie algebra of strictly upper
triangular matrices, $P_0$ is the projection onto the Lie subalgebra
of diagonal matrices and $P_-\equiv P^-$. Using the explicit form
of the invariant pairing (bilinear form on $gl((\infty))$) it is
easy to see that $P_{\pm}^*=P_{\mp}$,  $P_{0}^*=P_{0}$, and,
hence:
$$\frac{1}{2}(R-R^*)=P_+ -P_-.$$
As it follows from the results of  \cite{GM} (see also
\cite{SkrSIGMA})  if
$\mathfrak{g}=\mathfrak{g}_++
\mathfrak{g}_0+ \mathfrak{g}_-$ is a
triangular decomposition of a Lie algebra
$\mathfrak{g}$, i.e. $\mathfrak{g}_{\pm}$,
$\mathfrak{g}_0$ are closed Lie subalgebras and
$\mathfrak{g}_0$-modules, then any operator of the
form: $P_++R_0 -P_-$, where $P_{\pm}$ are projection operators
onto $\mathfrak{g}_{\pm}$, is a solution of modified
classical Yang-Baxter equation on $\mathfrak{g}$ if
$R_0$ is a solution of the modified classical Yang-Baxter equation
on $\mathfrak{g}_0$. Moreover, if the subalgebra
$\mathfrak{g}_0$ is abelian then  any operator $R_0$
(including trivial one) is a solution of the modified Yang-Baxter
equation on $\mathfrak{g}_0$. That is why operator
$P_+ -P_-$ is in this case a particular solution of the modified
Yang-Baxter equation on $\mathfrak{g}$.

On the other hand it is clear that the decomposition of the
algebra $\mathfrak{g}=gl((\infty))$ into the strictly
upper-triangular and strictly lower-triangular and diagonal part is a
triangular decomposition with abelian (diagonal) part
$\mathfrak{g}_0$. Hence for the considered
$R$-operator $$\frac{1}{2}(R-R^*)=P_+ -P_-$$ is indeed a solution
of mYBE on $gl((\infty))$ and  by the virtue of the Theorem
\ref{dus}  and Corollary \ref{cus} there indeed exist quadratic
and cubic Poisson structures on $gl((\infty))$ and its double.

Proposition is proven.

\paragraph{Acknowledgements}The second named author expresses  gratitude to
Guido Carlet for discussions. The work of the first named author is  partially supported by the European Research Council Advanced Grant \emph{FroM-PDE}.

\end{document}